\documentclass[11pt]{article}

\usepackage{epsfig,graphicx}
\usepackage{amsmath,amsfonts,amssymb}
\usepackage{cite}
\usepackage[usenames,dvipsnames]{color}
\usepackage{hyperref} 
\usepackage{breakurl}

\usepackage[normalem]{ulem}

\newcommand{\unit}{\leavevmode\hbox{\small1\kern-3.6pt\normalsize1}}

\def \GeV{{\mathrm{GeV}}}
\def \TeV{{\mathrm{TeV}}}

\def\lsim{\raise0.3ex\hbox{$\;<$\kern-0.75em\raise-1.1ex\hbox{$\sim\;$}}}
\def\gsim{\raise0.3ex\hbox{$\;>$\kern-0.75em\raise-1.1ex\hbox{$\sim\;$}}}

\allowdisplaybreaks

\def\neut{{\tilde{\chi}}^0_1}
\def\charg{{\tilde{\chi}}^{\pm}_1}
\def\relic{\Omega_{\tilde{\chi}^0_1} h^2}

\newcommand{\captions}{\sf\caption}

\newcommand{\EWft}{\Delta^{\rm (EW)}}
\newcommand{\mstopav}{\overline{m}_{\tilde{t}}}
\newcommand{\DMft}{\Delta^{\rm (DM)}}
\newcommand{\mstauR}{m_{\tilde{\tau}_1}}
\newcommand{\mhexp}{m_h^{(\rm exp)}}
\newcommand{\totft}{\Delta^{\rm (tot)}}

\begin{document}

\thispagestyle{empty}
\begin{flushright}
  FTUAM-16-44\\
  IFT-UAM/CSIC-16-134\\

\end{flushright}

\begin{center}
  {\Large \textbf{Low-mass neutralino dark matter in supergravity scenarios: phenomenology and naturalness
  } }  
  
  \vspace{0.5cm}
 M.~Peir\'o and S.~Robles
  \\[0.2cm] 
    
  {\textit{  Departamento de F\'{\i}sica Te\'{o}rica,
      and 
      Instituto de F\'{\i}sica Te\'{o}rica
      UAM-CSIC, \\[0pt] 
      Universidad Aut\'{o}noma de Madrid, 
      Cantoblanco, E-28049
      Madrid, Spain\\[0pt]  }}
  
\vspace*{0.7cm}
\begin{abstract}
The latest experimental results from the LHC and dark matter (DM) searches suggest that the parameter space allowed in supersymmetric theories is subject to strong reductions. 
These bounds are especially constraining for scenarios entailing light DM particles. Previous studies have shown that light neutralino DM in the Minimal Supersymmetric Standard Model (MSSM), 
with parameters defined at the electroweak scale, is still viable when the low energy spectrum of the model features light sleptons, in which case, the relic density constraint can be fulfilled.
In view of this, we have investigated the viability of light neutralinos as DM candidates in the MSSM, with parameters defined at the grand unification scale. 
We have analysed the optimal choices of
non-universalities in the soft supersymmetry-breaking parameters for both, gauginos and scalars, in order to avoid the stringent experimental constraints. 
We show that light neutralinos, with a mass as low as 25 GeV, are viable in supergravity scenarios if the gaugino mass parameters at high energy are very non universal, 
while the scalar masses can remain of the same order. These scenarios typically predict a very small cross section of neutralinos off protons and neutrons, thereby being very challenging for direct detection experiments. 
However, a potential detection of smuons and selectrons at the LHC, together with a hypothetical discovery of a gamma-ray signal from neutralino annihilations in dwarf spheroidal galaxies could shed light on 
this kind of solutions. 
Finally, we have investigated the naturalness of these scenarios, taking into account all the potential sources of tuning. 
Besides the electroweak fine-tuning, we have found that the tuning to reproduce the correct DM relic abundance and that to match the measured Higgs mass 
can also be important when estimating the total degree of naturalness.
\end{abstract}
\end{center}

	\newpage
	


\section{Introduction}

Weakly-interacting massive particles (WIMPs) are a very appealing kind of candidates to solve the dark matter (DM) problem, since generally they would be thermally produced in the correct amount in the early Universe. 
Among them, light WIMPs have received much attention in the last years in view of some direct~\cite{Bernabei:2003za,Bernabei:2008yi,Aalseth:2010vx,Angloher:2011uu,Aalseth:2011wp,Agnese:2013rvf,Aalseth:2014jpa} and indirect~\cite{Vitale:2009hr,Goodenough:2009gk,Hooper:2010mq,Hooper:2011ti,Abazajian:2012pn,Gordon:2013vta,Macias:2013vya,Abazajian:2014fta,Daylan:2014rsa,Murgia:2014,Calore:2014xka,Calore:2014nla,Calore:2015nua,Porter:2015uaa,Linden:2016rcf} detection experiments which might be seeing hints pointing towards a light DM particle.

However, these potential signals are being challenged by the null observations of other experiments. 
Several collaborations for DM direct detection have been able to place important constraints generally excluding the DM interpretation of these signals under certain assumptions. 
The most stringent bounds come from LUX \cite{Akerib:2013tjd,Akerib:2015rjg,Akerib:2016lao,Akerib:2016vxi}, PandaX-II \cite{Tan:2016zwf}, XENON \cite{Angle:2011th,Aprile:2012nq,Aprile:2013doa}, CDMS \cite{Ahmed:2009zw,Agnese:2013jaa}, SIMPLE \cite{Felizardo:2011uw}, KIMS \cite{Kim:2012rza}, CRESST~\cite{Angloher:2015ewa}, a combination of CDMS and EDELWEISS data \cite{Ahmed:2011gh} and SuperCDMS~\cite{Agnese:2014aze,Agnese:2015nto}. Furthermore, indirect detection experiments also provide an important constraint for low mass DM. 
Probably, the most important constraint for light WIMPs comes from the non detection of dwarf spheroidal galaxies (dSphs) in gamma rays by the Fermi-LAT collaboration~\cite{Ackermann:2013yva,Ackermann:2015zua}.  Although, under certain assumptions on the Milky Way DM profile, the Galactic centre provides a very stringent constraint as well~\cite{Gomez-Vargas:2013bea}. 
The dSph limits exclude WIMPs with a thermal cross section ($3\times10^{-26}$~cm$^3$/s) up to masses around 100 GeV. Moreover, the DM interpretation of the potential hints from the Galactic centre~\cite{Vitale:2009hr,Goodenough:2009gk,Hooper:2010mq,Hooper:2011ti,Abazajian:2012pn,Gordon:2013vta,Macias:2013vya,Abazajian:2014fta,Daylan:2014rsa,Murgia:2014,Calore:2014xka,Calore:2014nla,Calore:2015nua,Porter:2015uaa,Linden:2016rcf} are seriously challenged by this analysis.

On the theoretical side, there have been efforts in constructing well motivated theoretical models to explain the direct detection potential signals. In this context, neutralino DM in Supersymmetric theories is one of the best motivated candidates. Light neutralino DM has been studied in the most appealing realisations of Supersymmetry (SUSY) at the Electroweak (EW) scale such as the Minimal Supersymmetric Standard Model (MSSM)~\cite{Bottino:2002ry,Fornengo:2010mk,Vasquez:2010ru,Bottino:2011xv,Vasquez:2011yq,Arbey:2012na,Belanger:2012jn,Boehm:2013gst,Pierce:2013rda,Belanger:2013pna,Hagiwara:2013qya,Calibbi:2013poa}, the Next-to-MSSM (NMSSM)~\cite{Gunion:2005rw,Vasquez:2010ru,Das:2010ww,Cao:2011re,Carena:2011jy,Vasquez:2011yq,AlbornozVasquez:2011js,LopezFogliani:2012yq,Kozaczuk:2013spa,Han:2014nba,Cerdeno:2015jca}, 
and some minimal extensions of the NMSSM \cite{Cerdeno:2015jca,Cerdeno:2011qv,Cerdeno:2014cda,Cerdeno:2015ega}. Furthermore, some regions of the parameter space of these models are still viable in light of the most recent experimental constraints \cite{Pierce:2013rda,Belanger:2013pna,Hagiwara:2013qya,Calibbi:2013poa,Kozaczuk:2013spa,Han:2014nba,Cerdeno:2014cda,Cerdeno:2015ega,Cerdeno:2015jca}.

Nevertheless, an important question that still remains unanswered is whether or not these \textit{effective} Supersymmetric models can have a viable origin from a Supergravity (SUGRA) theory defined at the Grand Unification (GUT) scale. 
This is an interesting question since a conventional way of understanding the source of the soft supersymmetry-breaking terms is the breaking of Supergravity at a high energy scale. Moreover, by identifying the structure of the boundary conditions at the GUT scale,  
we could understand and quantify the degree of non-universality that is needed for having viable light neutralino DM, and hence providing a more accurate idea about the naturalness of these scenarios. 
The aim of this work is thus analysing if solutions with low-mass neutralinos in the MSSM can be achieved from a theory defined 
at the GUT scale and how natural these solutions are.
We show that obtaining light neutralinos which are viable dark matter candidates is actually possible from the point of view of SUGRA theories with soft terms defined at the GUT scale. 
Furthermore, we have determined the important role of non-universalities in the structures of the soft parameters and 
discussed the possibility of discovering these solutions by means of the LHC and indirect detection experiments. 

Another important issue that concerns MSSM scenarios is naturalness. Since the original motivation for SUSY models 
was precisely to avoid the huge fine-tuning associated with the hierarchy problem, we have performed a precise calculation of all the potential sources of fine-tuning in the MSSM. 
We have considered not only the contribution arising from the electroweak symmetry breaking condition, but also those required to reproduce different experimental and observational results 
such as the Higgs mass and the DM relic abundance. 
Finally, we have estimated the total amount of fine-tuning present in those solutions allowed by the experimental data.

This paper is organised as follows. 
In Section\,\ref{sec:eff}, we summarise the conditions under which very
light neutralinos can be viable dark matter candidates in the
MSSM with parameters defined at the EW scale. Then, in Section\,\ref{sec:sugra}, we investigate whether these
conditions can also be obtained from a high-energy description of the
theory. We analyse a general Supergravity model and investigate the
choices of non-universal gaugino and scalar masses that can produce
viable neutralinos with very light masses without violating any
experimental constraint. We also study the phenomenology of these scenarios in the context of
direct and indirect DM searches. Besides, in this section we study the constraints coming from the LHC data, namely, from searches involving direct EW production of charginos, selectrons and smuons. In Section\,\ref{sec:natur}, we summarise and calculate each of the contributions to the fine-tuning present in these scenarios, as well as the total amount of fine-tuning.
Finally, the conclusions are left for Section\,\ref{sec:concl}.

\section{Low-mass neutralinos in the effective MSSM}
\label{sec:eff}

Several studies~\cite{Bottino:2002ry,Vasquez:2010ru,Fornengo:2010mk,Vasquez:2011yq,Bottino:2011xv,Arbey:2012na,Belanger:2012jn,Pierce:2013rda,Belanger:2013pna,Scopel:2013bba,Hagiwara:2013qya,Calibbi:2013poa} have shown that a low-mass neutralino in the MSSM is a viable, albeit very fine-tuned, DM candidate. 
All the analyses performed after the Higgs boson discovery at the LHC consider the soft supersymmetry-breaking terms as input parameters defined at the SUSY scale. This is a framework often referred to as {\em effective} MSSM~\cite{Bottino:2002ry}, which exhibits a large flexibility since it does not incorporate any correlation among these parameters (although simplifying arguments are usually made). 

In order for the neutralinos, $\tilde{\chi}^0_1$, to be viable DM candidates they have to reproduce the correct value for their thermal abundance, $\relic$, which is actually non trivial in the MSSM, in particular when they are light. 
In the MSSM, within the regions of the parameter space where neutralinos are lighter than 50 GeV, there exist three dominant ways in which light neutralinos can annihilate efficiently. The first annihilation mechanism involves the exchange of CP-odd Higgses, the second involves the exchange of a $Z$ boson and, finally, the correct relic density can also be achieved through the $t$-channel exchange of sleptons. 

For neutralinos lighter than approximately 25 GeV, the first mechanism is the most efficient one and the relic density condition is fulfilled if $m_{A^0}$ is around $100-150$ GeV and $\tan\beta$ is $6-14$~\cite{Bottino:2002ry,Fornengo:2010mk,Bottino:2011xv}. Since the CP-odd Higgs mass and $\tan\beta$ parameters control the mass scale of the Higgs sector, the requirement of such light $m_{A^0}$ pushes the entire Higgs sector masses around this value. 
This scenario, known as \textit{intense coupling regime} \cite{Boos:2002ze}, is very restricted by ATLAS and CMS searches for neutral and charged scalars decaying into $\tau$-leptons \cite{CMS:gya,Aad:2012tj}. These unsuccessful searches have allowed to place stringent constraints on this scenario, almost ruling out this possibility. 
As previously mentioned, another possibility for neutralinos to account for the observed relic abundance occurs when they annihilate through the exchange of the $Z$ boson while satisfying the resonant condition, $m_{\tilde{\chi}^0_1}\approx M_Z/2$. 
However, this possibility restricts the mass of neutralinos to be in a few GeV range around the resonance and thus, it does not allow to obtain neutralinos far below this region. 

The annihilation through the exchange of sleptons, $\tilde{l}$, occurs in a $t$-channel diagram with a leptonic final state. This process is sufficiently efficient when the mediator is light enough. In Ref.~\cite{Vasquez:2011yq}, this option was investigated concluding that the sleptons must lie just above the LEP limit, $\mathcal{O}(90)$~GeV, in order for the annihilation to be high, and hence, to fulfil the relic density constraint. In this scenario, the requirement of a relatively light CP-odd Higgs boson is not longer necessary which makes easier to evade all the collider constraints~\cite{Vasquez:2010ru}, and to yield neutralinos as light as 15 GeV \cite{Belanger:2012jn,Pierce:2013rda,Belanger:2013pna}. 
Other works have pointed out the possibility that neutralinos could also be as light as 6 GeV if one sbottom is very light, with a mass splitting between them of a few GeV profiting from coannihilations\footnote{This scenario was motivated by the potential hints seen in direct detection experiments. Such a light sbottom increases considerably the elastic scattering cross section of neutralinos off protons and neutrons.} \cite{Arbey:2012na}. 
In view of the foregoing, hereafter we will focus only on this scenario.

The slepton exchange mechanism benefits especially from the presence of light staus, namely, their right-handed (RH) component~\cite{Belanger:2012jn}. 
This occurs specifically, due to an enhancement of the coupling of bino-like neutralinos to this component with respect to that of the left-handed (LH) component. The neutralino coupling to staus can be written as follows~\cite{Rosiek:1995kg,Pierce:2013rda}:
\begin{equation}
\begin{gathered}
g_{\tilde{\chi}^0_1\tilde{\tau}_1\tau_{LH}}=\sqrt{\frac{2}{v_u^2+v_d^2}}\left(m_Z\cos\theta_\tau(N_{11}s_W+N_{12}c_W)-N_{13}m_\tau\frac{\sin\theta_\tau}{\cos\beta}\right),\\
g_{\tilde{\chi}^0_1\tilde{\tau}_1\tau_{RH}}=-\sqrt{\frac{2}{v_u^2+v_d^2}}\left(2m_Z\sin\theta_\tau N_{11}s_W-N_{13}m_\tau\frac{\cos\theta_\tau}{\cos\beta}\right),
\label{eq:stau_neutralino_mssm}
\end{gathered}
\end{equation}
where $N_{1i}$ is the $i$-th component of the lightest neutralino in the basis $(\widetilde{B}, \widetilde{W}^3, \widetilde{H}^0_d, \widetilde{H}^0_u)$, $v_{u,d}$ are the vacuum expectation values (VEVs) of the up and down Higgses, respectively, and finally $\cos\theta_\tau$ is the stau mixing.
Since light neutralinos have to be mostly bino-like, $N_{11}\approx1$, in order to avoid the constraints from the $Z$ invisible decay, the coupling to the lightest RH stau in this case is higher than the corresponding coupling to the LH component. Therefore, the annihilation cross section of neutralinos through this mechanism will be increased due to the presence of a RH stau with a mass around 90 GeV, as close as possible to its lower experimental bound.

The most stringent limits on slepton masses come from LEP \cite{LEPSUSYWG}. These sparticles were constrained to have masses $m_{\tilde{e}}>$~100 GeV, $m_{\tilde{\mu}}>$~99 GeV, $m_{\tilde{\tau}_1}>$~80.5 GeV and $m_{\tilde{\nu}}>$~43 GeV. 
Notice that the actual limits depend on the neutralino mass, which is assumed to be the LSP to extract these bounds. Thus, we will incorporate this dependence in our results. 
Furthermore, using LHC data is also possible to constrain the mass of the sleptons, namely through the searches for direct production of 
selectrons and smuons \cite{Aad:2014vma,Khachatryan:2014qwa}.

Other constraints on SUSY particles might also affect the scenario considered along this paper. 
For the first two generation of squarks, inclusive searches constrain their mass to be above $608$ GeV, which holds for a compressed scenario~\cite{Aaboud:2016tnv}. 
For specific scenarios such as mass degeneracy in these families, this constraint can be as high as $1.5$~TeV~\cite{ATLAS:2017}. 
Consequently, in our analysis we will impose that the first two generation of squarks must be heavier than $1.5$ TeV, which is a conservative choice. 
For the third generation of squarks and the gluino masses, no restriction will be implemented, since as we will see in Section~\ref{sec:LHC}, we will analyse these cases separately.

Finally, low-energy observables are known to constrain severely SUSY theories, and namely, they can play an important role in this scenario. 
We will impose the recent measurement of the branching ratio of the $B_s\to \mu^+\mu^-$ process by the LHCb~\cite{Aaij:2013aka} and CMS~\cite{Chatrchyan:2013bka} collaborations, which collectively yields
$1.5\times 10^{-9}< {\rm BR}(B_s\to \mu^+\mu^-)< 4.3\times 10^{-9}$ at
95\% CL~\cite{Galanti-talk}. This branching ratio is strongly dependent on the Higgs sector of the model, and more specifically, on the CP-odd Higgs mass and $\tan\beta$. In SUSY theories the contributions to the flavour changing decay process $b\rightarrow s\gamma$ can be very important. We will take the experimental measurement of this observable at 2$\sigma$, which requires the range
$2.89\times 10^{-4}< {\rm BR}(b\rightarrow s\gamma)< 4.21\times
10^{-4}$, which takes into account
theoretical and experimental uncertainties added in quadrature \cite{Ciuchini:1998xy,D'Ambrosio:2002ex,Misiak:2006zs,
Misiak:2006ab,Amhis:2012bh}. It is also known that there exists a discrepancy between the measured values of the muon anomalous magnetic moment, $a_{\mu}$, and the SM prediction, that can be interpreted as a hint of SUSY. 
Although, we have not included this observable as a constraint in our analysis, we will comment on the prediction of $a_{\mu}$ for the solutions found in this work.

\section{Low-mass neutralinos in SUGRA theories}
\label{sec:sugra}

As usual, the SUGRA model presented in this paper is defined in terms of the soft super\-symmetry-breaking parameters, which comprise
mass parameters for the scalars and gauginos, as well as trilinear parameters associated with the Yukawa couplings.    
Successful Radiative Electroweak Symmetry-Breaking (REWSB) is achieved by imposing the
following boundary condition on the $\mu$ parameter at the EW scale,
\begin{equation}
  \mu^2 = \frac{m_{H_d}^2 - m_{H_u}^2 \tan^2 \beta}{\tan^2 \beta -1 } - 
  \frac{1}{2} M_Z^2\,,
  \label{eq:electroweak}
\end{equation} 
in terms of the Higgs soft mass parameters. The ratio of the Higgs VEVs,
$\tan\beta$, will hereafter be considered as a free parameter.

In general, the soft SUSY breaking terms arising from a large class of string scenarios,
namely symmetric orbifold constructions, present a certain degree of non universality~\cite{deCarlos:1992pd,Brignole:1993dj,Baer:2000gf}. Furthermore, within some realisation of type I string models is possible to obtain non universal gaugino masses, $A$-terms and scalar masses~\cite{Khalil:2000ci}. This motivates us to consider a general scenario in which some of the soft terms can be different from each other. 
Moreover, it is known that in the minimal SUGRA scenario of the MSSM light neutralinos ($m_{\tilde{\chi}^0_1}\lesssim 50$~GeV) are not allowed in light of experimental constraints such as the chargino mass constraint and the Higgs mass measurement~\cite{Ellis:2012aa,Baer:2012uya}. 
Due to this fact, the first crucial step to find scenarios comprising light neutralinos is to allow departures from the universal scenario in the gaugino sector which allow the lightest neutralino to be dominated by the bino component. 
Bino-like neutralinos are obtained by lowering $|M_1|$ with respect to $|M_2|$ and $|\mu|$. 
These two parameters, $M_2$ and $\mu$, are involved in the chargino mass matrix, and thus, LEP constraint rules out the region of the parameter space in which $|M_2|$, $|\mu|\lesssim100$~GeV. 
From this condition, it follows that bino-like neutralinos require $|M_1|\lesssim100$~GeV and then $m_{\tilde{\chi}_1^0}\simeq |M_1|$ at the EW scale. 
We will keep a universal relation between $M_2$ and $M_3$ at the GUT scale. 
In what follows, to differentiate between parameters evaluated at the GUT and EW scales, unless otherwise specified, we denote the latter with the upper index EW. 

Let us start defining the soft masses and trilinear parameters at the GUT scale. 
We will consider a gaugino sector parametrized by
 \begin{equation}
  M_1,\,M_2=M_3\,.
  \label{eq:gauginopara}
\end{equation}

Regarding the scalar sector, we will allow departures from universality in the Higgs soft masses, which in light of Eq.~\eqref{eq:electroweak} control the $\mu$ parameter at the EW scale. 
This has a profound impact on the phenomenology of the solutions at the EW scale since $\mu$ determines the Higgsino component of the lightest neutralino and therefore the DM phenomenology~\cite{Cerdeno:2004zj}. 
In consequence, we have used as input parameters, 
 \begin{equation}
  m_{H_{d}},\,m_{H_{u}}\,.
  \label{eq:higgspara}
\end{equation}

\begin{table}
  \begin{center}
    \begin{tabular}{|c|c|}
      \hline
      Parameter & Range\\
      \hline
      \hline 
      $M_1$& $[-110,110]$\\
      $M_2=M_3$&  $[-1000,1000]$\\
      $\tan\beta$ & $[1.5,60]$\\
      $A_U$& $[-7000,7000]$\\
      $A_D$& $0$\\
      $A_E$& $[-7000,7000]$\\
      $m_{H_u}$& $[1,7000]$\\
      $m_{H_d}$& $[1,7000]$\\
      $m_{\tilde{L}_2}=m_{\tilde{L}_1}$& $[0,7000]$\\    
      $m_{\tilde{L}_3}$& $[0,7000]$\\    
      $m_{\tilde{E}_2}=m_{\tilde{E}_1}$& $[0,7000]$\\   
      $m_{\tilde{E}_3}$& $[1,7000]$\\  
      $m_{\tilde{Q}_{1,2,3}}=m_{\tilde{U}_{1,2,3}}=m_{\tilde{D}_{1,2,3}}$& $[0,7000]$\\  
      \hline 
    \end{tabular}
    \caption{Input parameters for the scan defined at the GUT scale. Masses and trilinear parameters are given in GeV.}
    \label{tab:scan}
  \end{center}
\end{table} 

As mentioned in the previous section, the relic density constraint requires slepton masses close to the LEP limit, namely, a RH stau mass around 90 GeV at the EW scale. 
For this reason, we have considered non-universalities in the slepton soft masses, which at high energy are described by the following parameters, 
 \begin{equation}
  m_{\tilde{L}_3},\,m_{\tilde{E}_3},\,m_{\tilde{L}_2}=m_{\tilde{L}_1},\,m_{\tilde{E}_2}=m_{\tilde{E}_1}\,.
  \label{eq:sleptonpara}
\end{equation}
Since the LHC limits on the slepton masses are especially stringent for the first and second generations, 
we have taken the soft masses of these two generations as degenerated, while the third generation parameters are free to vary independently. 

For the squarks soft masses, we have assumed universality so that they are given at the GUT scale by just one free parameter, 
 \begin{equation}
  m_{\tilde{Q}_{1,2,3}}=m_{\tilde{U}_{1,2,3}}=m_{\tilde{D}_{1,2,3}}=m_{\tilde{Q}}\,.
  \label{eq:squarkpara}
\end{equation}

Finally, the three trilinear parameters $A_U$, $A_D$ and $A_E$ are considered family independent. 
Particularly important is the top trilinear, $A_U$, which controls the Higgs mass and affects very strongly the ${\rm BR}(B_s\to \mu^+\mu^-)$~\cite{Baek:2005wi}. 
The slepton trilinear term, $A_E$, might play an important role through the stau mixing. 
For the down type squarks, we have chosen $A_D=0$. This parameter controls the mixing of the down sector and might modify radiative corrections to the Higgs mass at large $\tan\beta$, which would result in an increase of the Higgs mass\cite{Carena:2011aa}.
This could be important to achieve a Higgs mass around 125 GeV. However, the large $\tan\beta$ regime is disfavoured by flavour constraints, and thus we would expect that this choice does not have an impact on the scenarios investigated throughout this paper\footnote{Notice that some flavour constraints such as ${\rm BR}(B_s\to \mu^+\mu^-)$ scale as $1/m_{A^0}^4$ and hence could be also avoided for large pseudoscalar masses.}. 
Therefore, we have considered the following trilinear parameters as inputs, 
 \begin{equation}
  A_{U},\,A_{D}=0,\,A_E\,.
  \label{trilinearpara}
\end{equation}

\begin{figure}[t!]
  \begin{center}
    \epsfig{file=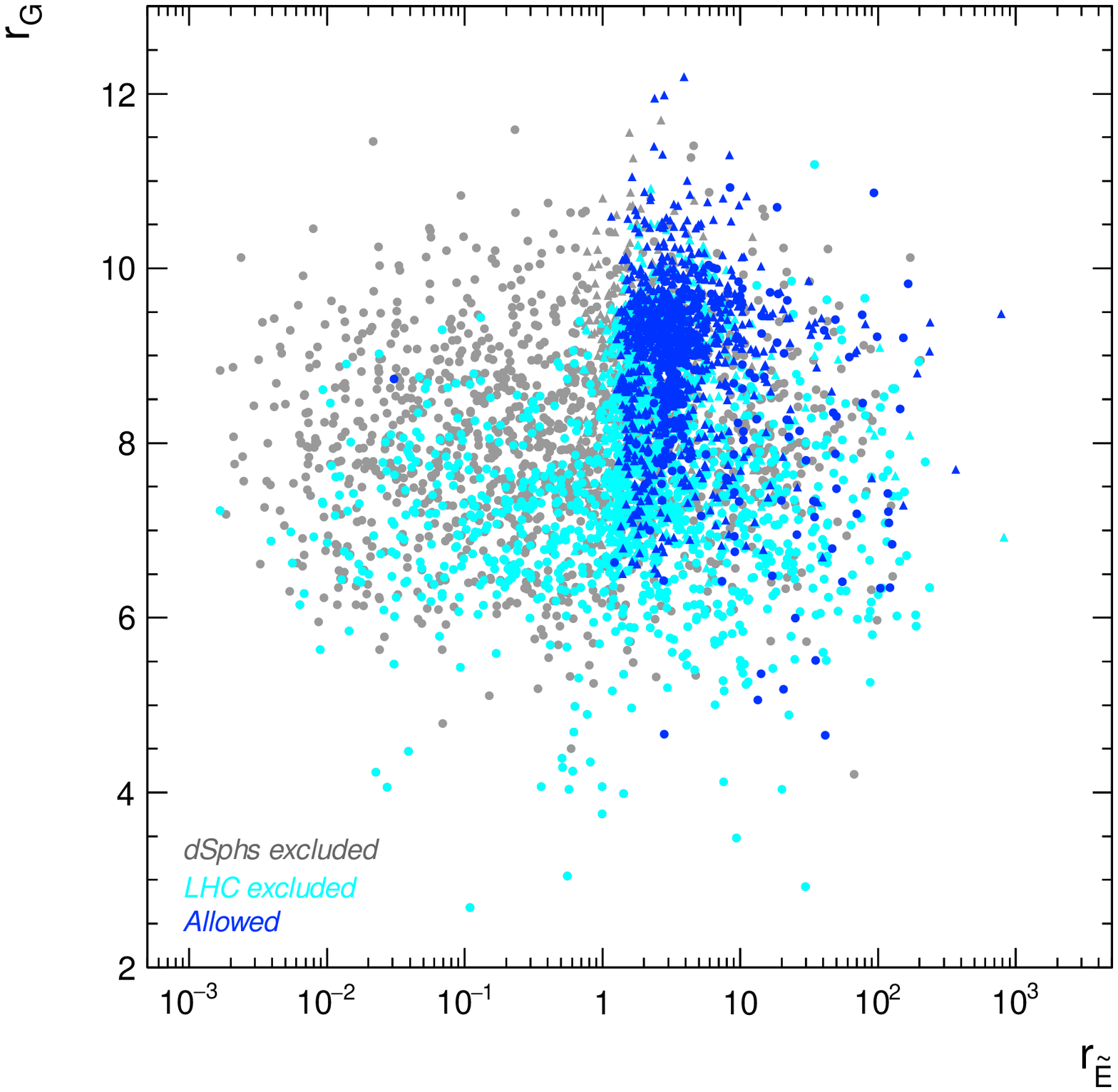,width=7.6cm}
   \epsfig{file=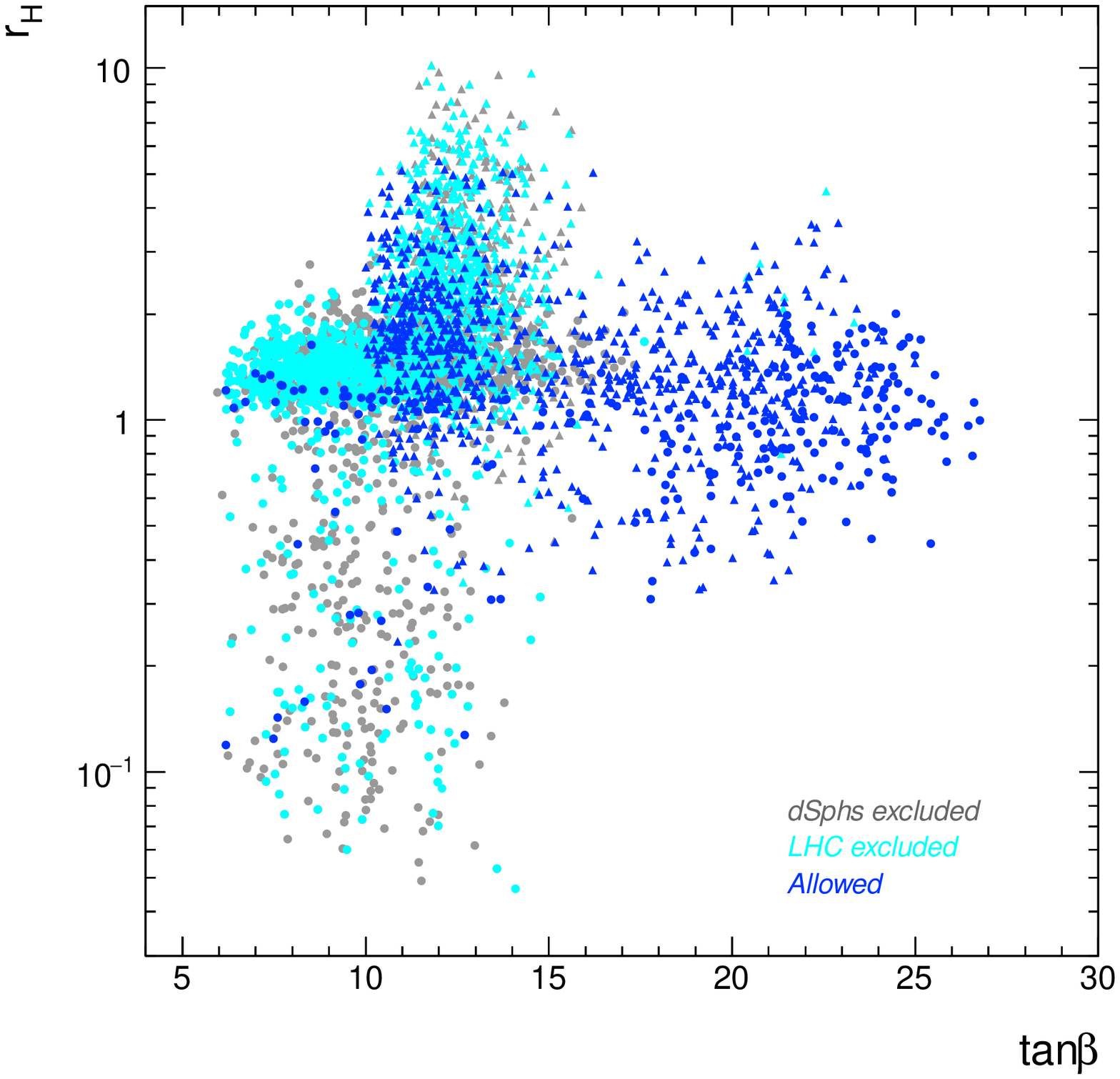,width=7.6cm}
  \end{center}
    \captions{Universality patterns in the scalar and gaugino sectors. 
  All points fulfil the experimental constraints from the Higgs sector, LEP limits on new particles, relic density and direct detection searches.
  Points excluded by dSph bounds for a $\tau^+\tau^-$ final state are shown in grey, solutions excluded by LHC searches for sparticles in cyan and 
  points fulfilling all the experimental constraints are shown in blue. Circle points ($\bigcirc$) correspond to solutions with $M_1>0$, $M_2>0$ and $A_U<0$, whereas triangles ($\bigtriangleup$) represent those with $M_1<0$, $M_2<0$ and $A_U>0$.}
  \label{fig:inputs}
\end{figure}

Following these criteria, we have performed a scan over the MSSM
parameter space where the aforesaid input parameters are varied
according to Table\,\ref{tab:scan}. In order to efficiently explore the 12-th dimensional parameter space considered, we have used {\tt MultiNest 3.10} \cite{Feroz:2007kg,Feroz:2008xx,Feroz:2013hea}. 
To that end, we have built a likelihood function, whose 
parameters are the neutralino relic density consistent with latest Planck results~\cite{Ade:2013zuv}, $m_{H^0_1}$, ${\rm BR}(B_s\to \mu^+\mu^-)$, and ${\rm BR}(b\to s\gamma)$, calculated with  {\tt micrOMEGAs 4.3} \cite{Belanger:2013oya,Barducci:2016pcb}, 
and taken as Gaussian probability distribution functions around the measured values with 
$2\sigma$ deviations.  This likelihood function is 
used to generate MCMC and to find regions of the 
parameter space that maximise the likelihood. Using {\tt MultiNest} allows 
us to scan the parameter space of the model more efficiently, since relatively 
few evaluations are needed to converge to regions of maximum likelihood. 
Let us clarify at this point that a statistical approach to the problem presented here is out of the scope of this work. Our aim is to provide an answer to whether or not light neutralinos can be obtained from a general SUGRA scenario, not how statistically favoured are these scenarios in light of the current experimental data. Other constrains such as those coming from direct and indirect DM searches have been included once the solutions have been found and thus, they are not used to generate the MCMC.  
To evolve from the GUT scale down to the EW scale the input parameters, we have used {\tt SoftSUSY 3.7.2}~\cite{Allanach:2001kg}. 
All collider constraints, including those from LEP, Tevatron and the LHC, have been implemented using the interfaces available in the  {\tt micrOMEGAs 4.3} code to different tools~\cite{Barducci:2016pcb}. 
More specifically, the constraints on the Higgs sector were implemented through the codes {\tt HiggsBounds} \cite{Bechtle:2008jh,Bechtle:2013wla}, {\tt HiggsSignals}~\cite{Bechtle:2013xfa} and {\tt Lilith}~\cite{Bernon:2015hsa}, 
LEP limits were imposed through the functions provided by {\tt micrOMEGAs 4.3}, and finally LHC constraints on SUSY particles have been calculated with {\tt SModelS}~\cite{Kraml:2013mwa,Kraml:2014sna}.

For convenience, let us define the ratios,
\begin{equation}
r_{G}\equiv\frac{M_2}{M_1},\hspace{2cm}r_{H}\equiv\frac{m_{H_d}}{m_{H_u}},\hspace{2cm}r_{\tilde{E}}\equiv\frac{m_{\tilde{E_3}}}{m_{\tilde{E_2}}},
\end{equation}
which measure the departure from universality in the gaugino, Higgs and slepton sectors, respectively. 
In Figure~\ref{fig:inputs}, we show the values of these ratios for all the solutions found fulfilling the experimental constraints from the Higgs sector, LEP limits on new particles, relic density and direct DM searches. 
Points excluded by Pass~8 data from dSphs are displayed in grey, those excluded by LHC searches for SUSY particles are shown in cyan and the solutions that fulfil all the experimental constrains in blue. Finally, circular shaped points ($\bigcirc$) correspond to solutions in which $M_1>0$, $M_2>0$ and $A_U<0$, whereas triangle shaped points ($\bigtriangleup$) represent those with $M_1<0$, $M_2<0$ and $A_U>0$.

\begin{figure}[t!]
  \begin{center}
    \epsfig{file=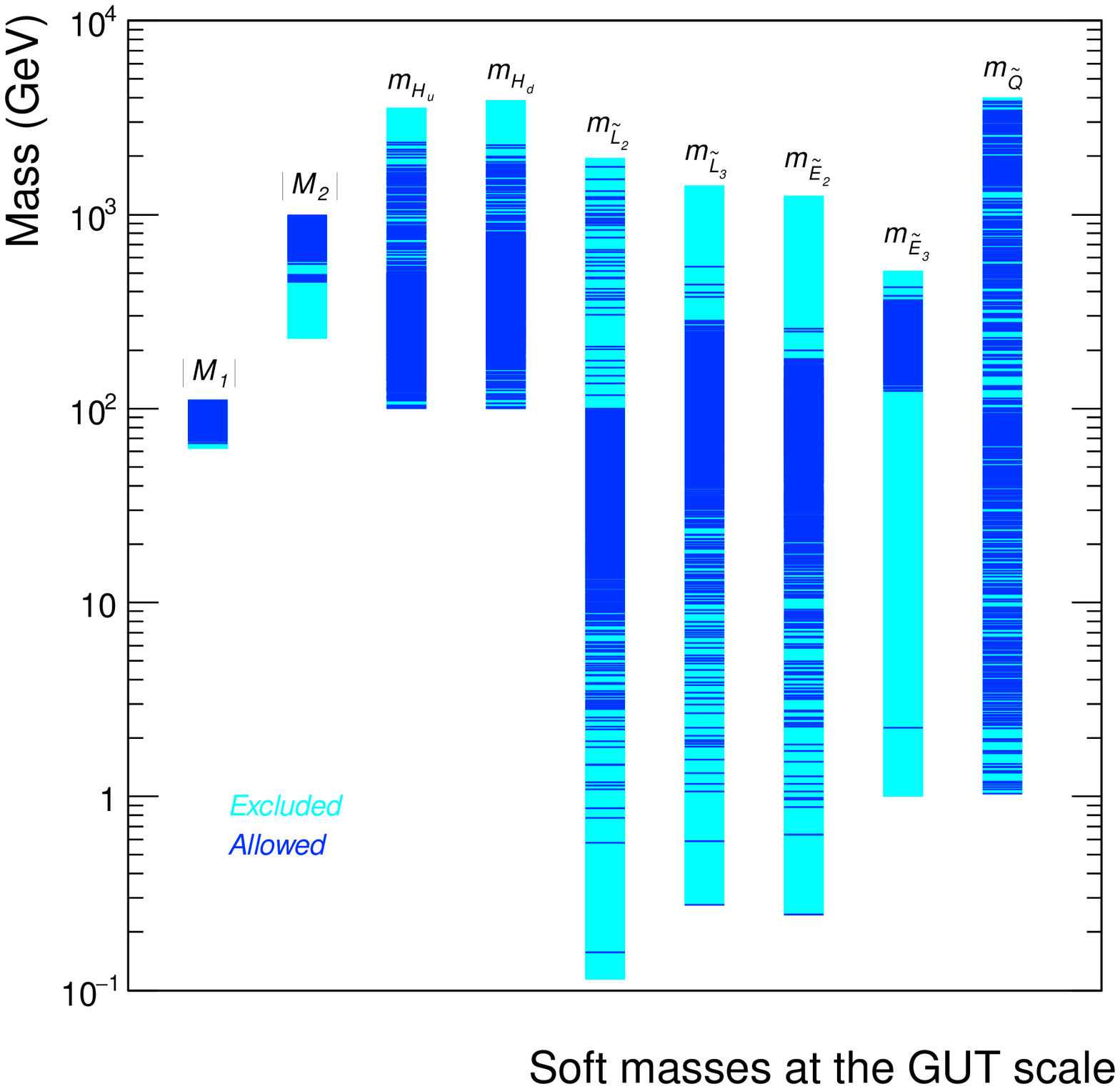,width=7.6cm}
  \end{center}
    \captions{Spectrum of the soft masses at the GUT scale.  Regions shaded in blue correspond to the points fulfilling all the experimental constrains, solutions excluded by LHC searches for sparticles and dSph bounds are shaded in cyan. }
  \label{fig:specGUT}
\end{figure}

In the left panel of Figure~\ref{fig:inputs}, we have plotted $r_{G}$ versus $r_{\tilde{E}}$. As it can be seen, all viable points exhibit a precise relation between gaugino masses, $2.5\lesssim r_{G}\lesssim 12$. 
As stated above, the chargino mass is proportional to $M_2^{EW}$, and it is restricted by LEP null searches which results in $r_{G}>1$ (the universal value). 
On top of that, our choice $M_2=M_3$ yields $M_2^{EW}\simeq3 M_3^{EW}$, and thus the lower bound on the gluino mass forbids $M_2$ below $\sim 250$ GeV (see Figure ~\ref{fig:specGUT}). 
All this together is translated into values of $r_{G}\gtrsim2.5$. On the other hand, the ratio of slepton soft masses, $r_{\tilde{E}}$, is much less constrained. 
The relic density constraint, which prefers a light RH stau, favours $m_{\tilde{E_3}}$ to be below the LEP constraint\footnote{The running of this parameter is positive \cite{Martin:1997ns}.} 
and hence it is found close to the lower edge of the scan range (see Figure ~\ref{fig:specGUT}). 
On top of this, the values of $m_{\tilde{E_2}}=m_{\tilde{E_1}}$ are affected by the ATLAS bound on the first and second generation of sleptons which clearly generates $r_{\tilde{E}}<1$. 
However, since we are allowing $A_E$ to be a free parameter\footnote{Remind that this parameter is taken as family independent.}, 
the slepton masses can be deeply influenced by it, rendering into a less constrained $r_{\tilde{E}}$ ratio.

In the right panel of Figure~\ref{fig:inputs}, $r_{H}$ is plotted versus $\tan\beta$, with the same colour code as the left panel. In this plane, we can see that the relation between the Higgs soft
masses is, in general, surprisingly close to the universal value. The values of  $\tan\beta$ found are in the range 6-27, which is mainly a consequence of the Higgs mass bound because in this range the maximal mixing scenario is reached. 

Let us comment that when universal relations between the soft parameters at the GUT scale are taken, it is has been pointed out that the SM-like Higgs mass requires $|A_U/m_{\tilde{Q}}|>2$ \cite{Aparicio:2012iw}. Unlike this, we have obtained points with the correct Higgs mass with $|A_U/m_{\tilde{Q}}|<2$ due to the flexibility provided by assuming non universality. 
For the slepton sector, we have found that in general our solutions accumulate around $|A_E/m_{\tilde{E}_3}|\approx5$, but they can reach very high values when the soft mass $m_{\tilde{E}_3}<10$~GeV and $|A_E|$ is at the TeV scale.

\begin{figure}[t!]
  \begin{center}
     \epsfig{file=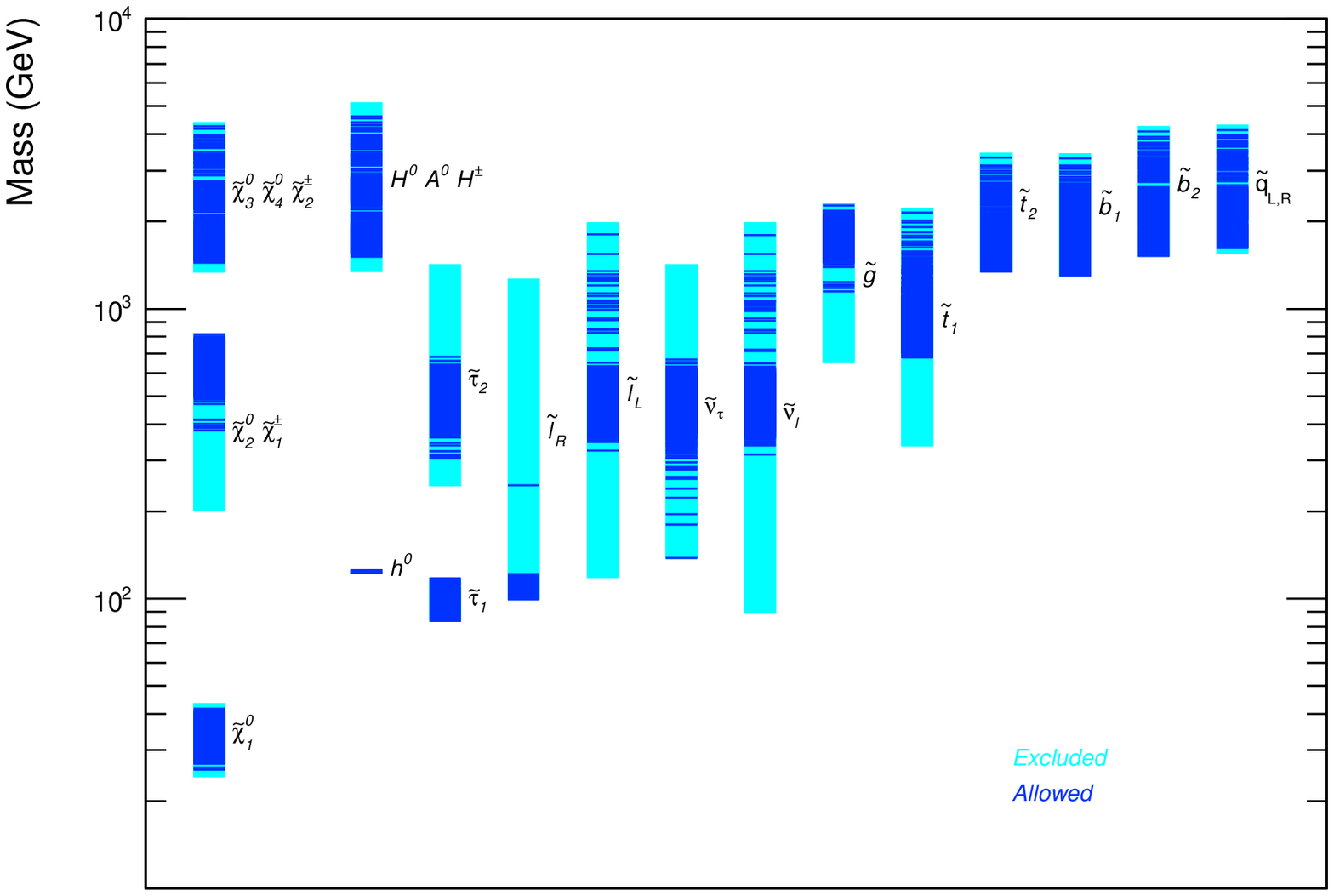,width=10.0cm}
  \end{center}
    \captions{Spectrum of physical masses. Colour code as in Figure~\ref{fig:specGUT}. }
  \label{fig:specEW}
\end{figure}

In Figure~\ref{fig:specGUT}, the spectrum of soft masses at the GUT scale is shown. Notably, we observe that low-mass neutralinos require a definite type of spectrum. Most of the allowed solutions comprise soft masses at the GUT scale in the range of $10$~GeV and $1$~TeV, being lighter, in general, the soft masses corresponding to the slepton sector. 
This translates, at the EW scale, into spectra that can be taken as representative of these scenarios, as it can be observed in Figure~\ref{fig:specEW}.
More specifically, 
these scenarios comprise a gap between the EW sector, represented by the sleptons and the lightest neutralino, and the coloured sector which is sited at the TeV scale as a consequence of the LHC null searches. As we will see later, this kind of scenarios can be explored at the LHC by means of searches involving EW direct production.

Finally, the presence of relatively light smuons, charginos and neutralinos, as well as large values of $\tan\beta$, produce sizeable SUSY contributions to the muon anomalous magnetic moment. 
From $e^+e^-$ data, the SUSY contribution to this observable is constrained to be $10.1\times10^{-10} < a^{SUSY}_{\mu} <42.1\times10^{-10}$ at 2$\sigma$. 
However, tau data favour a slightly smaller discrepancy, $2.9\times10^{-10} < a^{SUSY}_{\mu} <36.1\times10^{-10}$ at 2$\sigma$~\cite{Davier:2010nc}, while a more recent update using the Hidden Local Symmetry model leads to $16.5\times10^{-10} < a^{SUSY}_{\mu} <48.6\times10^{-10}$ at 2$\sigma$~\cite{Benayoun:2012wc}. 
We have checked that most of the solutions found satisfying all the experimental constraints cluster around $a^{SUSY}_{\mu} \approx20\times10^{-10}$, in great agreement with all the mentioned data sets. 
Only a subset of these solutions fall well below $10^{-10}$, but these are still within the 2$\sigma$ lower bounds. This is a consequence that arises from the presence of a light EW sector in the allowed solutions (see Figure~\ref{fig:specEW}). 

\subsection{Direct detection}

Direct DM searches are based on the elastic scattering of DM off nuclei inside a detector. For any WIMP candidate, the WIMP-nucleus elastic cross section depends, at the microscopic level, on the WIMP-quark interaction strength.
\begin{figure}[t!]
\begin{center}
   \epsfig{file=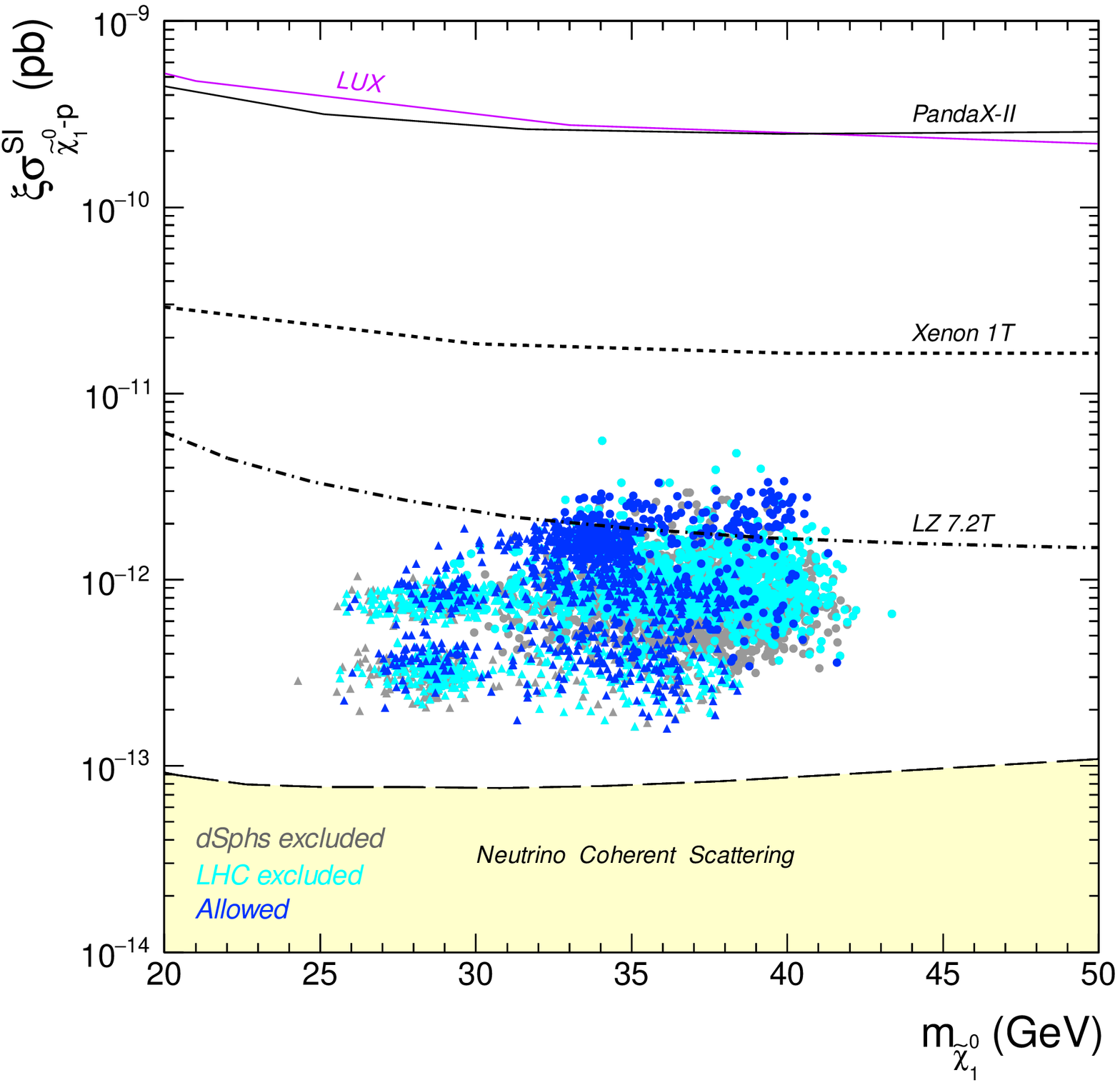,width=7.6cm}
    \epsfig{file=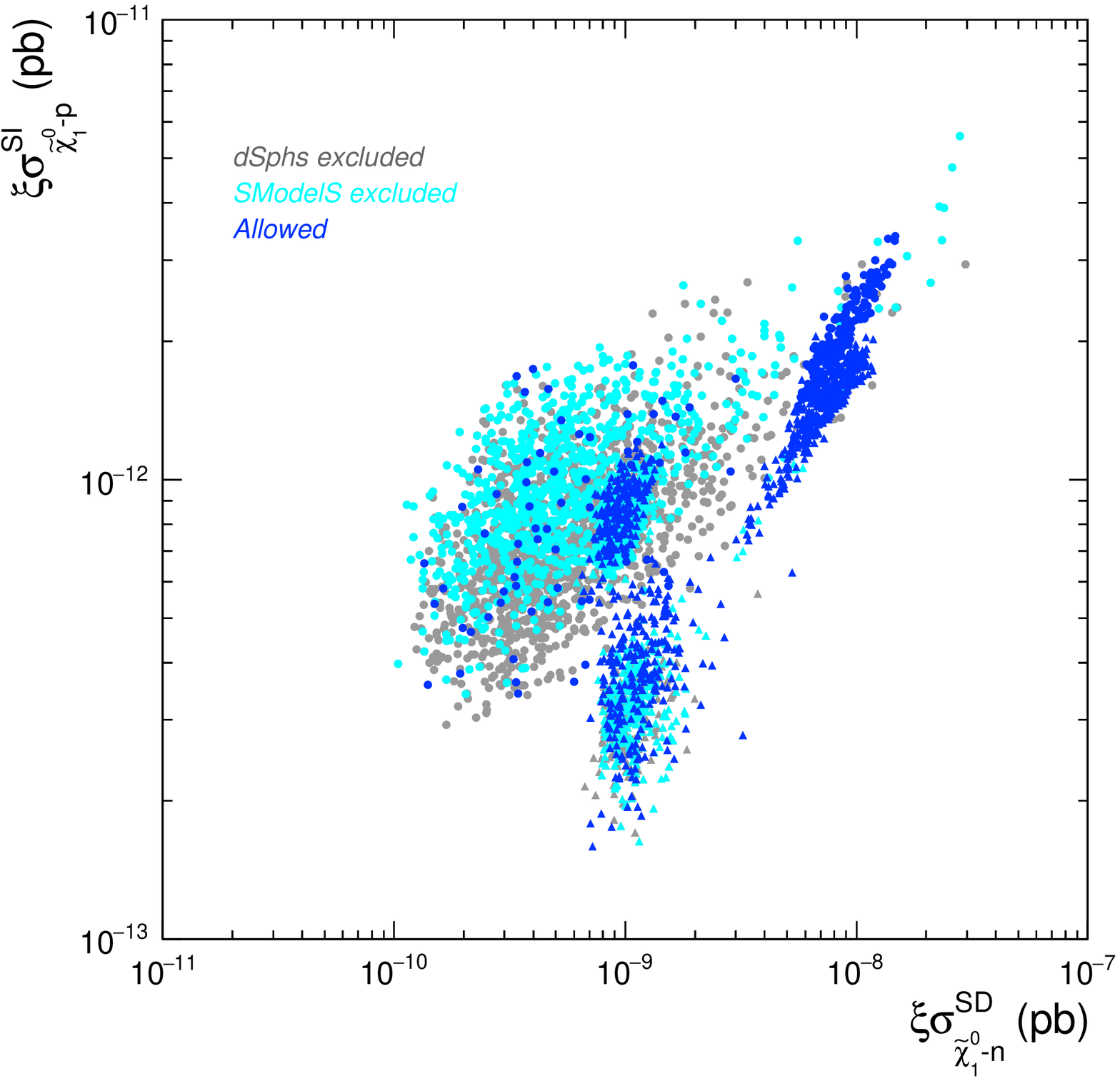,width=7.6cm}		
\captions{Left: Theoretical predictions for $\xi\sigma_{\tilde{\chi}_1^0-p}^{SI}$ as a function of the neutralino mass using the same colour code as in Figure \ref{fig:inputs}. 
As a reference, we have included the current constraints from LUX (solid violet line) and PandaX-II (solid black line), 
and the future prospects of LZ(dot-dashed line) and Xenon 1T (dashed line). 
In addition, the irreducible neutrino background is shown as a yellow region at the bottom of the plot \cite{Billard:2013qya}. Right: Spin independent cross section off protons, $\xi\sigma_{\tilde{\chi}_1^0-p}^{SI}$, as a function of the spin dependent cross section off neutrons, $\xi\sigma_{\tilde{\chi}_1^0-n}^{SD}$. The colour code is as in the left panel. }
\label{fig:dd_mssm}
\end{center}
\end{figure}
For the MSSM neutralino (and in general for any Majorana fermion, neglecting momentum and velocity suppressed operators, the effective Lagrangian describing this interaction reads
\begin{equation}
L_{eff}=\sum_{q_i}\alpha_{q_i}\bar{\chi}\chi\bar{q_i}q_i+\xi_{q_i}\bar{\chi}\gamma_5\gamma_\mu\chi\bar{q_i}\gamma_5\gamma^\mu q_i\,,
\label{eq:mssm_effdd}
\end{equation}
where the sum runs over the six quarks, and the coefficients $\alpha_{q_i}$ and $\xi_{q_i}$ can be found in Refs. \cite{Jungman:1995df,Belanger:2008sj}. 
The first term in Eq. (\ref{eq:mssm_effdd}) corresponds to the scalar interactions, which contribute to the spin-independent (SI) interactions, 
and the latter, the axial-vector interactions, contribute to the spin-dependent (SD) interactions. 

For the SI interactions, the neutralino-nucleon cross section can be written as
\begin{equation}
\sigma_{p,n}^{SI}=\frac{4\mu_{p,n}^2}{\pi}f_{p,n}^2\,,
\end{equation}
where $\mu_{p,n}$ is the neutralino-nucleon reduced mass and $p,n$ stand for protons and neutrons, respectively\footnote{Notice that the same can be done for the SD interactions. For instance see Ref.~\cite{Cerdeno:2013gqa}.}. These parameters can be further decomposed as, 
\begin{equation}
\frac{f_{p,n}}{m_{p,n}}=\sum_{q_i} f_{q_i}^{p,n}\frac{\alpha_{q_i}}{m_{q_i}}\,,
\end{equation}
where $m_{q_i}$ is the corresponding quark mass. The parameter $\alpha_{q_i}$ represents the effective coupling of neutralinos to quarks, and it must be calculated for the elastic scattering of neutralinos off quarks mediated by CP-even Higgs bosons and squarks~\cite{Jungman:1995df}. 
The $f_{q_i}^{p,n}$ terms parametrize the quark content of the nucleon for either protons and neutrons. These quantities depend on the light quark mass ratios ($m_u/m_d$ and $m_s/m_d$), the pion sigma term $\sigma_{\pi N}$, and the operator $\sigma_s=m_s\langle p|\bar{s}s|p\rangle$~\cite{Belanger:2008sj}. 
For our calculations, we have used the values from the latest version of the {\tt micrOMEGAs} code~\cite{Barducci:2016pcb}. Nevertheless, it is important to remark that these parameters are extracted using lattice QCD calculations, and hence, are subject to important uncertainties that can affect the elastic scattering cross section~\cite{Ellis:2008hf}. 
It is also worth mentioning that for SD interactions the uncertainties related to the calculation of the SD structure functions can lead to important differences in the theoretical predictions of 
the differential event rate of the elastic scattering of neutralinos off a nucleus as well~\cite{Cerdeno:2012ix}.

In Figure \ref{fig:dd_mssm} (left panel), the SI elastic scattering cross section of the lightest neutralino off protons, $\xi\sigma_{\tilde{\chi}_1^0-p}^{SI}$, is shown versus the lightest neutralino mass. 
The fractional density, $\xi=\min[1,\Omega_{\tilde{\chi}_1^0}h^2/0.11]$, is included to account for the reduction in the direct detection rate in the cases where the neutralino only contributes to a fraction of the total DM density, assuming that it is present in the DM halo in the same proportion as in the Universe\footnote{The use of $\xi$ in this case is not very important since most of the solutions found do not present a relic abundance below $0.1$.}. 
Unfortunately, from the experimental point of view, the SI cross sections predicted for these scenarios are remarkably small, even out of the reach of future experiments like Xenon 1T~\cite{Aprile:2015uzo} (dashed line), only LZ~\cite{Cushman:2013zza} (dot-dashed line) could probe a small fraction of our results. 
However, there would be a chance to detect them via direct detection experiments since these solutions are above the so-called irreducible neutrino background \cite{Billard:2013qya}. 
The bino-like nature of the lightest neutralino, $N_{11}^2\approx1$, resulting from $M_1\ll M_2,\mu$,  decreases the coupling to the scalar Higgs, which is proportional to the Higgsino components, $N_{13}$ and $N_{14}$. 
The remaining contribution to the cross section arises from the squark $s$-channel exchange. The cross section corresponding to this interaction is proportional to
\begin{equation}
\sigma_{\tilde{\chi}_1^0-p}^{SI}\propto\frac{|N_{11}|^4}{m_{\tilde{q}}^4}\,,
\end{equation}
and hence, the LHC bounds on the squark masses are translated into a strong decrease of this cross section. 

In Figure \ref{fig:dd_mssm} (right panel), we show the SI cross section off protons versus the SD cross section off neutrons\footnote{Xe-based experiments are mostly sensitive to the neutron component of the SD cross section.}. As it can be seen,  the SD cross section predictions found for these solutions are very small, 
and in general would not contribute to the differential event rate more than the SI contribution. 
The main contribution to the SD cross section comes from the t-channel $Z$ boson exchange, but the coupling of neutralinos to the $Z$ boson is proportional to the Higgsino mixing. Therefore, this cross section is suppressed as well.

It is worth noting that predictions for direct detection cross sections are a consequence of the REWSB boundary condition of Eq.~\eqref{eq:electroweak}. 
All the solutions experimentally allowed entail values of $\mu$ at the TeV scale, which along with the fact that low-mass neutralinos have a relatively low value of $M_1$, pose a difficult challenge for direct DM searches. 
This is, indeed, the main difference with solutions found in Refs.~\cite{Vasquez:2011yq,Pierce:2013rda} with soft parameters at the EW scale. 
The amount of the Higgsino component found in these solutions increases substantially the cross sections through the t-channel Higgs exchange.

\subsection{Indirect detection}

As already stated, the predictions of these scenarios for direct DM searches are strongly influenced by the $\mu$ parameter value, and hence are very low, out of the reach of the current experiments. 
Nonetheless, it is known that thermal relics generally predict annihilation cross sections in DM haloes that lie in the ballpark of the current searches, especially for light DM candidates. 
Since in the scenarios presented here neutralinos are thermally produced in the early Universe, indirect detection experiments might provide a hopeful window to probe these scenarios.

To estimate the thermally averaged cross section, usually an expansion in powers of $x\equiv T/m$ is employed. 
In this approximation, the annihilation cross section times the relative velocity can be written as $\langle\sigma v\rangle\simeq a+6bx$, which holds for non-relativistic particles at the freeze-out temperature as long as there are not $s$-channel resonances and thresholds for new final states. 
In the case studied here, this approximation can be safely used, at least to extract interesting features about expected values\footnote{Our results make use of the whole numerical calculation provided by {\tt micrOMEGAs}.}. 
In the limit of vanishing stau mixing, the $a$ and $b$ parameters can be written as follows~\cite{Pierce:2013rda}
\begin{equation}
\begin{gathered}
a=\frac{m_{\tilde{\chi}^0_1}^2}{8\pi}\left[\frac{g_{\tilde{\chi}^0_1\tilde{\tau}_1\tau_R}g_{\tilde{\chi}^0_1\tilde{\tau}_1\tau_L}}{(m_{\tilde{\tau}^0_1}^2+m_{\tilde{\chi}^0_1}^2)}\right]^2\,,\\
b\approx\frac{m_{\tilde{\chi}^0_1}^2}{48\pi}\left[\frac{(g_{\tilde{\chi}^0_1\tilde{\tau}_1\tau_R}^4+g_{\tilde{\chi}^0_1\tilde{\tau}_1\tau_L}^4)(m_{\tilde{\tau}^0_1}^4+m_{\tilde{\chi}^0_1}^4)}{(m_{\tilde{\tau}^0_1}^2+m_{\tilde{\chi}^0_1}^2)^4}\right]\,.
\label{eq:sv_mssm}
\end{gathered}
\end{equation}
On the one hand, in this limit the $a$ parameter, known as the $s$-wave contribution, is proportional to the Higgsino mixing, and thus it can be neglected. 
On the other hand, the $b$ parameter, known as the $p$-wave contribution, is going to dominate the cross section. 
However, a cross section dominated by the $p$-wave term is temperature suppressed, by a factor $T/m$. 
This means that, when the temperature decreases the annihilation cross section decreases as well. 
Therefore, the annihilation in DM haloes is suppressed relative to that of the early Universe and 
a lower gamma ray flux is expected (with respect to the canonical cross section, $3\times10^{-26}$cm$^3/$s).
\begin{figure}[t!]
\begin{center}
	    \epsfig{file=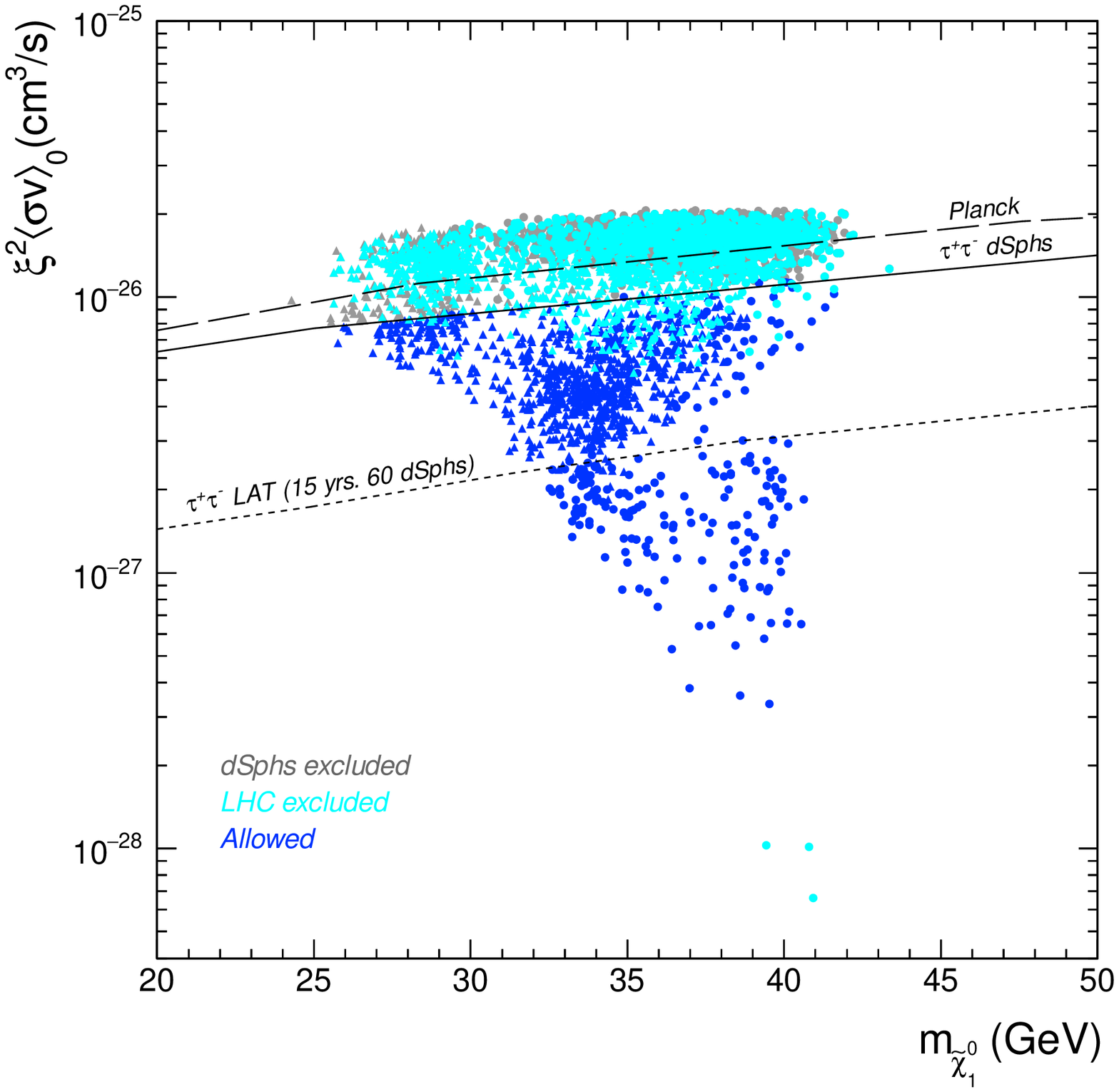,width=7.6cm}
\captions{Thermally averaged neutralino annihilation cross section in the galactic halo, $\xi^2\langle\sigma v\rangle_0$, as a 
function of the neutralino mass. The solid line corresponds to the upper bound on $\langle\sigma v\rangle$ derived from an analysis of dSph galaxies for pure $\tau^-\tau^+$ 
using the Pass~8 reprocessed data set. The dotted line corresponds to the expected sensitivity of the LAT experiment~\cite{Charles:2016pgz}, and 
the dashed line stands for the Planck limits~\cite{Ade:2015xua}. The colour code is as in Figure ~\ref{fig:inputs}.}
\label{fig:id_mssm}
\end{center}
\end{figure}

In Figure~\ref{fig:id_mssm}, the theoretical predictions for the thermally averaged cross section of neutralinos in the Galactic halo, $\xi^2\langle\sigma v\rangle_0$, as a function of its mass are depicted. 
The fractional density is included squared because the annihilations of neutralinos in the halo depend on its density squared. 
The main annihilation channel is driven by the exchange of a light RH stau which yields a pair of $\tau$-leptons in the final state. 
In order to compare our findings with current limits, we have also included those from dSph galaxies by Fermi-LAT collaboration (solid line) for a $\tau^+\tau^-$ final state \cite{Ackermann:2015zua} 
and the Planck bounds (dashed line)~\cite{Ade:2015xua}. 
Despite of the suppressed cross sections respect to the thermal value, we can observe that the recent Pass~8 data from Fermi-LAT collaboration is starting to probe these scenarios. 
Interestingly, the scenarios unconstrained by Pass~8 could be tested in the near future provided that the Fermi-LAT experiment accumulates more exposure~\cite{Charles:2016pgz} (dotted line). 
We see that the prospects are very promising for these scenarios. Most of the solutions corresponding to $M_1<0$ and $M_2<0$ (triangles) will be probed by the LAT experiment, and only a small subset of solutions, 
those with cross sections below $2\times10^{-27}$~cm$^3/$s approximately, will remain hidden. 

\subsection{LHC searches for SUSY particles}
\label{sec:LHC}

Many of the points entailing light neutralinos are ruled out by LHC searches for SUSY particles. 
Even though, some solutions are excluded by direct production of gluinos~\cite{Chatrchyan:2013iqa} and 
stops~\cite{ATLAS:2013cma}, the most restricting channels are those involving the electroweak production of sleptons, charginos and chargino-neutralino par production. 
Namely, two opposite sign leptons and missing transverse energy $E_{T}^{miss}$ final states through the processes 
$pp\rightarrow\tilde{\chi}^{\pm}\tilde{\chi}^{\mp}\rightarrow l\tilde{l}\nu\tilde{\nu}\rightarrow\tilde{\chi}^0_1\tilde{\chi}^0_1 l\nu$ and 
$pp\rightarrow\tilde{l}\tilde{l}\rightarrow\tilde{\chi}^0_1\tilde{\chi}^0_1 ll$ entailing chargino and slepton pair production, 
and three leptons and missing transverse energy $E_{T}^{miss}$ final states through the processes $pp\rightarrow\tilde{\chi}^{\pm}\tilde{\chi}^{0}_2\rightarrow l\tilde{l}l\tilde{\nu}\rightarrow\tilde{\chi}^0_1\tilde{\chi}^0_1 l\nu$ and $pp\rightarrow\tilde{\chi}^{\pm}\tilde{\chi}^{0}_2\rightarrow l\tilde{l}\nu\tilde{l}\rightarrow\tilde{\chi}^0_1\tilde{\chi}^0_1 ll$ involving chargino-neutralino production~\cite{Aad:2014vma,Khachatryan:2014qwa}.
The former being the most constraining of these two LHC searches.

\begin{figure}[t!]
  \begin{center}
    \epsfig{file=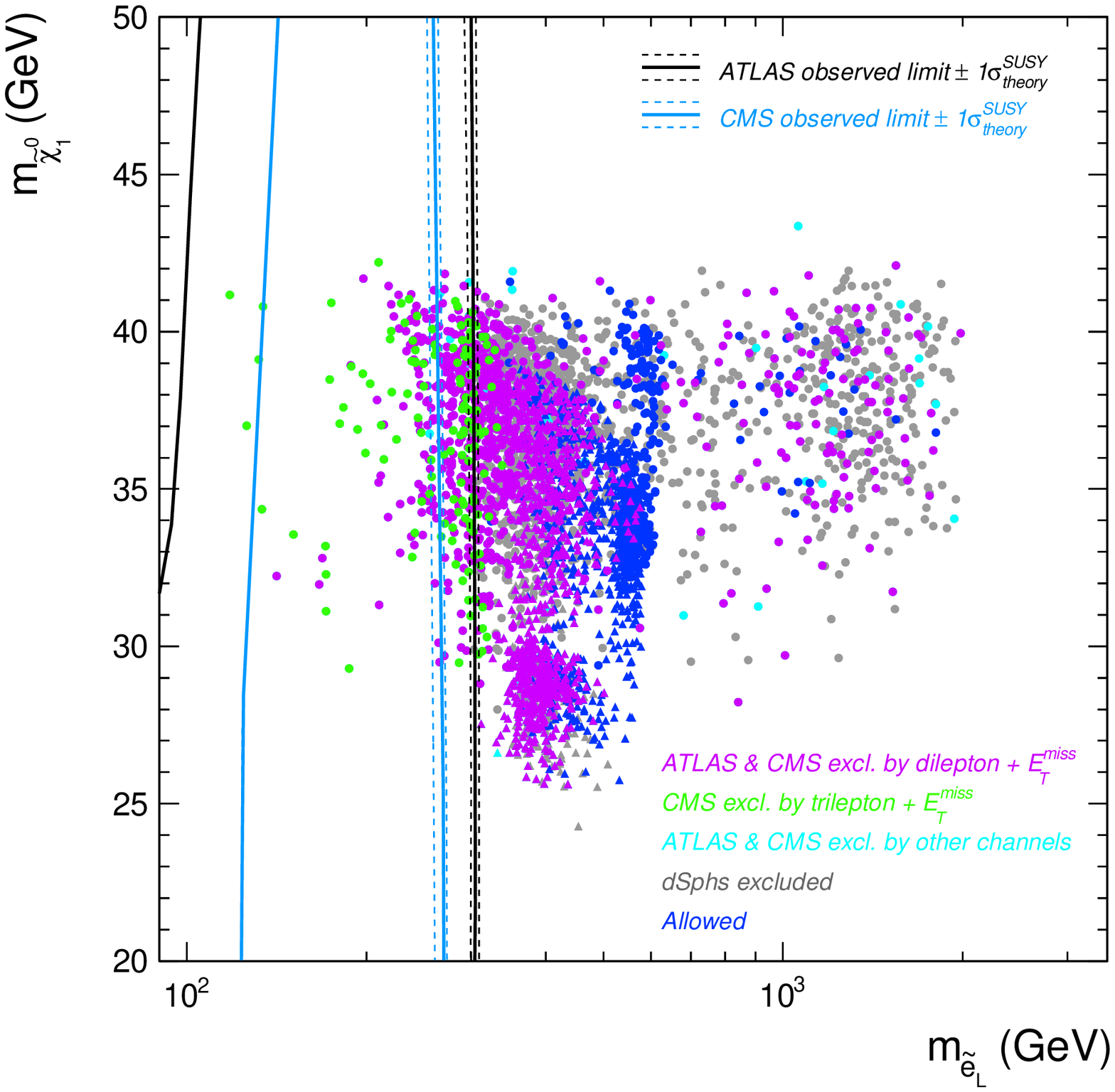,width=7.6cm}
    \epsfig{file=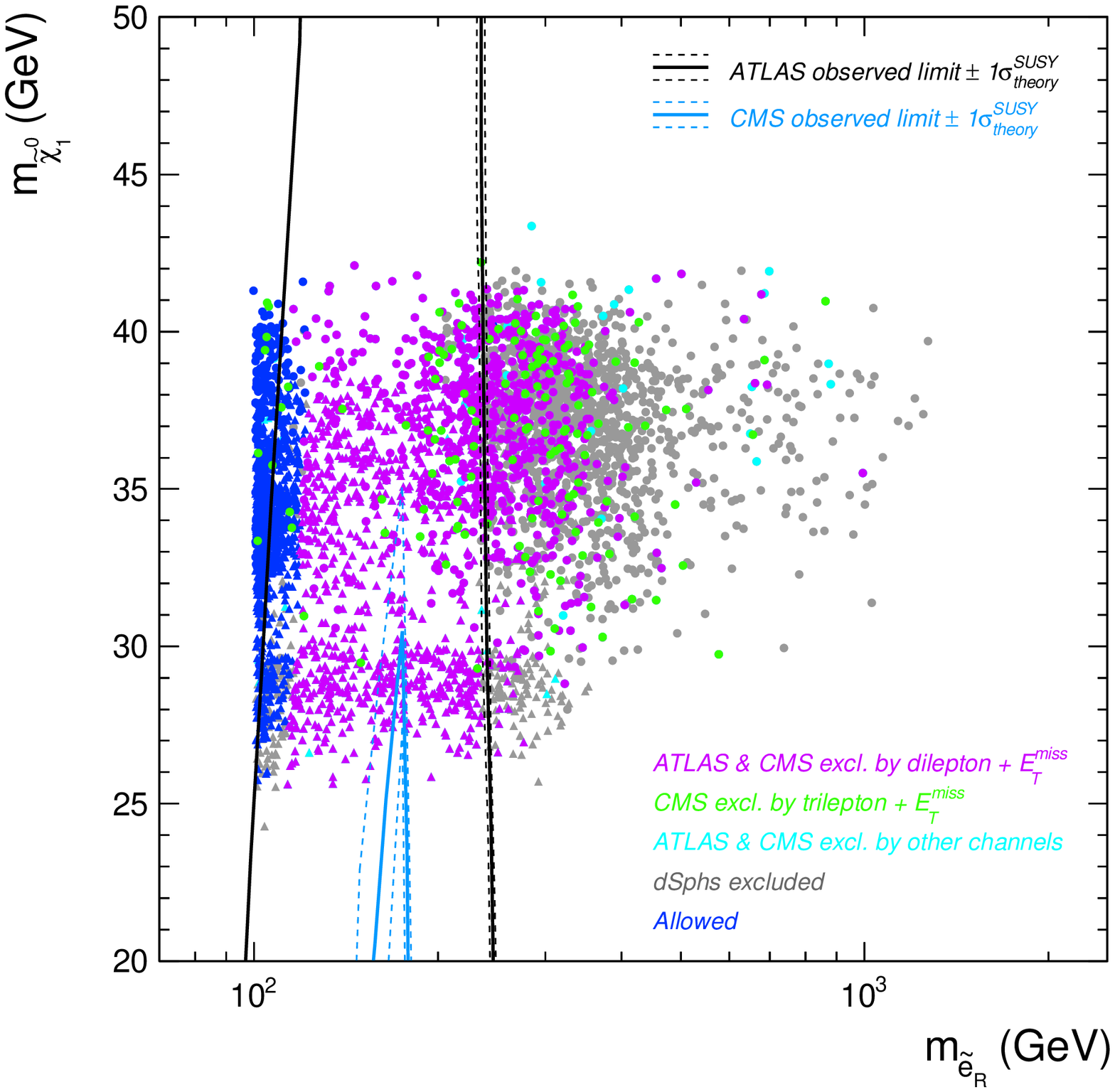,width=7.6cm}
    \epsfig{file=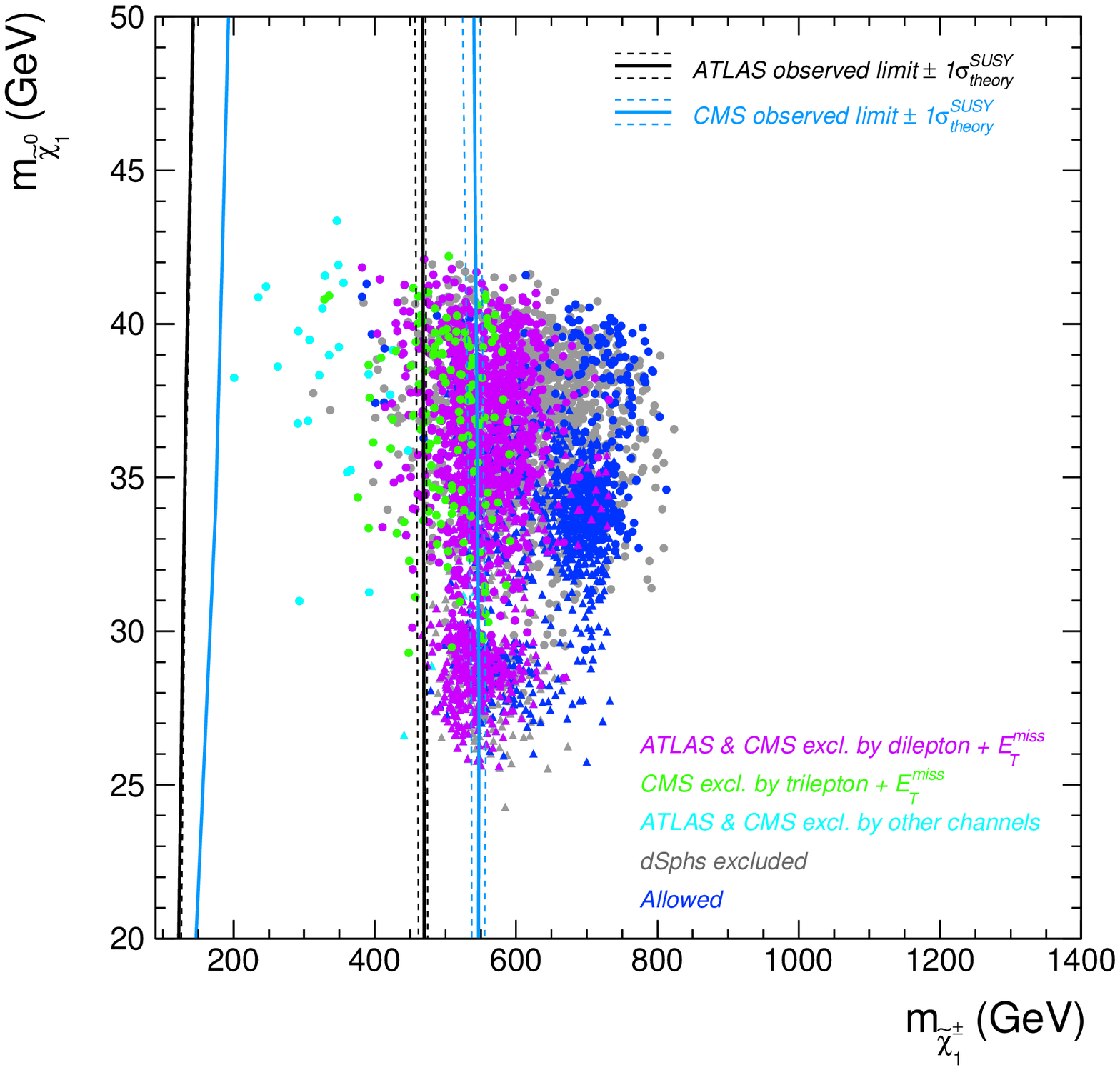,width=7.6cm}    
  \end{center}
    \captions{LHC search for direct slepton and chargino par production.
    Violet points denote solutions excluded by ATLAS and CMS dilepton searches, green points are ruled out by CMS trilepton searches, whereas cyan points are excluded by ATLAS and CMS  gluino and stop bounds. 
    Solutions excluded by Fermi-LAT bounds from dSphs are shown in grey and finally allowed solutions are depicted in blue.
    As a reference, we include the $95\%$ CL ATLAS (black lines) and CMS (light-blue lines) limits on pure left-handed slepton pair production 
    (upper left panel), right-handed slepton pair production (upper right panel) and chargino par production (bottom panel).  }
  \label{fig:LHC}
\end{figure}

As previously mentioned, we have calculated the LHC bounds from SUSY particle searches with {\tt SModelS}. 
In Figure~\ref{fig:LHC}, we show the solutions excluded by dilepton (violet) and trilepton (green) searches, whereas 
solutions ruled out by gluino and stop limits are displayed in cyan. Grey points are excluded by Femi-LAT dSph upper limits, 
illustrating the complementarity between indirect detection experiments and collider searches, and finally 
the solutions fulfilling all the experimental constraints are shown in blue. 
It is worth remarking that there are points ruled out by more than one channel, mainly by dilepton and trilepton final states, 
in those cases we have represented only the most constraining channel. 
As expected, most of our excluded results are ruled out by right-handed slepton pair production (see upper-right panel of Figure~\ref{fig:LHC}), 
in particular by ATLAS results.
Note that we have depicted the ATLAS (black lines) and CMS (light-blue lines) bounds at 95\% CL as a reference, 
since they depend on the production cross section,  and in the chargino production bound 
 $m_{\tilde{l}}=m_{\tilde{\nu}}=(m_{\neut}+m_{\charg})/2$ is also assumed. 

The latest ATLAS~\cite{ATLAS:2016uwq}  and CMS~\cite{CMS:2016gvu} results at 13~TeV on electroweak production of SUSY particles in multilepton 
final states have not being incorporated in our results, since they do not include slepton pair production, which is the most restrictive process 
for the scenarios analysed here. Other LHC bounds that can potentially probe part of our allowed solutions are the most recent 
searches for gluino pair production from ATLAS~\cite{ATLAS:2016uzr} and CMS~\cite{CMS:2016mwj}, and direct stop pair production from CMS~\cite{CMS:2016hxa}.

Regarding future colliders, currently, there are no prospects for improving the limits on the direct production of sleptons, charginos and 
chargino - neutralino at the high luminosity LHC (HL-LHC), when sleptons are not so heavy. These are the most constraining channels for the scenarios we have studied. 
Nonetheless, we expect that some of our solutions could be probed by direct gluino, top and sbottom pair production at the HL-LHC \cite{CMS:2013ega,ATL-PHYS-PUB-2013-011,ATL-PHYS-PUB-2014-010}. 
On the other hand, linear colliders such as the ILC and CLIC can shed light on the scenarios studied here by the direct production of 
sleptons, charginos and neutralinos, provided they have sufficiently large centre of mass energy, they might either measure very precisely 
their mass, spin and couplings or push harder the LEP constraints \cite{Linssen:2012hp,Baer:2013vqa,Moortgat-Picka:2015yla,CLIC:2016zwp}.

\section{Naturalness of SUGRA scenarios}
\label{sec:natur}

Being the original motivation for low-energy SUSY to solve the hierarchy problem, 
i.e. the delicate balance between the soft terms required to reproduce the smallness of the EW scale, 
it is interesting to check the degree of naturalness of our solutions in the SUGRA model considered here. 
In consequence, we must analyse all the potential sources of 
fine-tuning that could affect these scenarios in 
order to obtain a total estimation of the degree of fine-tuning of our allowed solutions. 

\subsection{Electroweak fine-tuning}

Let us start by the most important source of fine-tuning and, in fact, the most extensively studied in the literature, that induced to reproduce the electroweak scale. 
In what follows, we summarise the origin and the method to measure the EW fine-tuning in the MSSM, and then, we compute it for the solutions 
that fulfil all the experimental constraints discussed in the previous sections.

In the MSSM, the vacuum expectation value of the Higgs, $v^2/2 = |\langle H_u\rangle|^2 +|\langle H_d\rangle|^2$, is given, at tree-level, 
by the minimization relation given in Eq.~\eqref{eq:electroweak}. 
This expression can be rearranged as
\begin{equation}
-\frac{1}{8}(g^2+g'^2)v^2 = -\frac{M_Z^2}{2}=\mu^2-\frac{m_{H_d}^2 - m_{H_u}^2\tan^2\beta}{\tan^2\beta-1} \ .
\label{eq:min} 
\end{equation}
As is well known, the value of $\tan\beta$ must be moderately large, in such a way that the tree-level Higgs mass,  
$(m_h^2)_{\rm tree-level}=M_Z^2 \cos^22\beta$, is as large as possible. Otherwise, the radiative corrections needed to reconcile the Higgs mass with its experimental value, 
would imply large stop masses and thus, an extremely fine-tuned scenario. 

The absolute value of the terms on the  r.h.s. of  Eq.~\eqref{eq:min}  are typically much larger than $M_Z^2$, 
thereby giving rise to a potential fine-tuning associated with the electroweak symmetry breaking. 
The radiative corrections to the Higgs potential reduce the fine-tuning to a certain extent, due to the running of the effective quartic coupling 
of the SM-like Higgs from its initial value at the SUSY 
threshold,\footnote{A convenient choice of the SUSY-threshold is the average stop mass, 
since the one-loop correction to the Higgs potential is dominated by the stop contribution.} which hereafter we identify with the low-energy (LE) scale, 
$\lambda(Q_{\rm LE})=\lambda(Q_{\rm threshold})=\frac{1}{8}(g^2+g'^2)$, down to the value at the electroweak scale, $\lambda(Q_{EW})$. 
Essentially, this is equivalent to replace $M_Z^2 \rightarrow m_h^2$ in Eq.~\eqref{eq:min} (for more details, see Ref.~\cite{Casas:2014eca}), i.e.
\begin{equation}
\label{eq:minh}
-\frac{m_h^2}{2}=\mu^2(LE)-\frac{m_{H_d}^2(LE) - m_{H_u}^2(LE)\tan^2\beta}{\tan^2\beta-1} \ . 
\end{equation}
From this expression, we will calculate the electroweak fine-tuning. 
Notice that the terms on the r.h.s of Eq.~\eqref{eq:minh} have to be evaluated at the low-energy scale.

A common practice to quantify the EW fine-tuning is to use the parametrization first proposed by Ellis et al.~\cite{Ellis:1986yg} 
and Barbieri and Giudice~\cite{Barbieri:1987fn}, which for Eq.~\eqref{eq:minh} reads 
\begin{eqnarray}
\Delta_{\theta_i}^{\rm (EW)} &=& \frac{\theta_i}{ m_h^2} \frac{\partial m_h^2}{\partial \theta_i} = -2\frac{\theta_i}{m_h^2} \frac{\partial}{\partial \theta_i}\left( \mu^2(LE)-\frac{m_{H_d}^2(LE) - m_{H_u}^2(LE)\tan^2\beta}{\tan^2\beta-1} \right) \ , \nonumber \\
\EWft &\equiv& {\rm Max}\ \left|\Delta_{\theta_i}^{\rm (EW)}\right|\ ,
\label{eq:BG}
\end{eqnarray}
where $\theta_i$ is an independent parameter that defines the model under consideration and $\Delta_{\theta_i}^{\rm (EW)}$ is the fine-tuning associated with it. 
Typically, $\{\theta_{i}\}$ are the initial (high-energy) values of the soft terms and the $\mu$ parameter. 

It is worth emphasising that the value of $\EWft$ can be interpreted as the inverse of the $p$-value to obtain $m_h^2$ from Eq.~\eqref{eq:minh} equal or smaller than
the experimental value. If $\theta$ is the parameter that gives the maximum $\EWft$ and $\delta \theta$ represents the $\theta$-interval for which $m_h^2\lsim m_h^{2~(\rm exp)}$, then 
assuming that the  $\theta$ parameter has an approximate flat probability distribution, the $p$-value can be expressed as
\begin{equation}
\label{toy2}
p{\rm -value} \simeq \left|\frac{\delta \theta}{\theta_0}\right| \equiv \Delta^{-1}\ .
\end{equation}
We refer the reader to Refs.~\cite{Ciafaloni:1996zh,Casas:2014eca} for further details on the statistical meaning of $\EWft$.

In order to use the standard measure of the fine-tuning, Eq.~\eqref{eq:BG}, it is necessary to write the r.h.s of the minimization equation~\eqref{eq:minh} as a function of the initial (input) parameters of the model. 
This, in turn, implies to write the low-energy (LE)\footnote{We recall that the low-energy scale is the scale at 
which we set the SUSY threshold and the supersymmetric spectrum is computed, taken here as the average stop mass.} values of $\mu^2$, $m_{H_u}^2$ and $m_{H_d}^2$ in terms of the initial, 
GUT scale parameters, which are related through the renormalization group equations (RGEs). 
Fortunately, dimensional and analytical consistency dictates the form of such dependence,
\begin{eqnarray}
\label{mHu_gen_fit}
m_{H_u}^2(LE)&=&
c_{M_3^2}M_3^2 +c_{M_2^2}M_2^2 +c_{M_1^2}M_1^2 + c_{A_U^2}A_U^2+c_{A_UM_3}A_UM_3 +c_{M_3M_2}M_3M_2+ \cdots 
\nonumber\\
&&\cdots +c_{m_{H_u}^2}m_{H_u}^2 +c_{m_{Q_3}^2} m_{Q_3}^2 +c_{m_{U_3}^2}m_{U_3}^2+\cdots \ ,
\\
\label{mHd_gen_fit}
m_{H_d}^2(LE)&=&c_{m_{H_d}^2}m_{H_d}^2 +
 c_{M_2^2}M_2^2 + \cdots \ ,
\\
\mu(LE)&=&c_\mu \mu \ ,
\label{mu_gen_fit}
\end{eqnarray}
where the terms on the r.h.s are the soft terms at the GUT scale.
The numerical coefficients, $c_{M_3^2}, c_{M_2^2},...$ are obtained by fitting the result of the numerical integration of the RGEs 
to eqs.~(\ref{mHu_gen_fit}, \ref{mHd_gen_fit}, \ref{mu_gen_fit}), a task that was performed following the prescription described in Ref.~\cite{Casas:2014eca}, 
this is, considering the different threshold scales involved and two-loop RGEs. 
First, we have performed the integration of the SM and MSSM RGEs between the corresponding scales  
with {\tt SARAH 4.9.1} \cite{Staub:2013tta} (see Ref.~\cite{Casas:2014eca} for further details). Then, we have 
calculated the coefficients for a generic MSSM, i.e. a model with the following initial parameters  
\begin{equation}
\Theta_i=\left\{\mu, M_1,M_2,M_3,A_U,A_D,A_E, m_{H_u}^2, m_{H_d}^2, m_{U_3}^2, m_{Q_3}^2, \cdots\right\}. 
\end{equation}
Next, we have determined the coefficients of eqs.~\eqref{mHu_gen_fit} and \eqref{mHd_gen_fit} for the model analysed here, applying the relations among the initial parameters indicated in Section~\ref{sec:sugra}. 
Finally, we have evaluated $\EWft_{\theta_i}$ for our set of initial parameters, namely 
\begin{equation}
\Theta_i=\left\{\mu, M_1,M_2,A_U,A_E, m_{H_u}^2, m_{H_d}^2, m_{Q}^2, m_{L_2}^2, m_{L_3}^2, m_{E_2}^2, m_{E_3}^2\right\}, 
\end{equation}
by using eqs.~\eqref{mHu_gen_fit} - \eqref{mu_gen_fit} in \eqref{eq:BG} and determined $\EWft$. 

\begin{figure}[t!]
  \begin{center}
    \epsfig{file=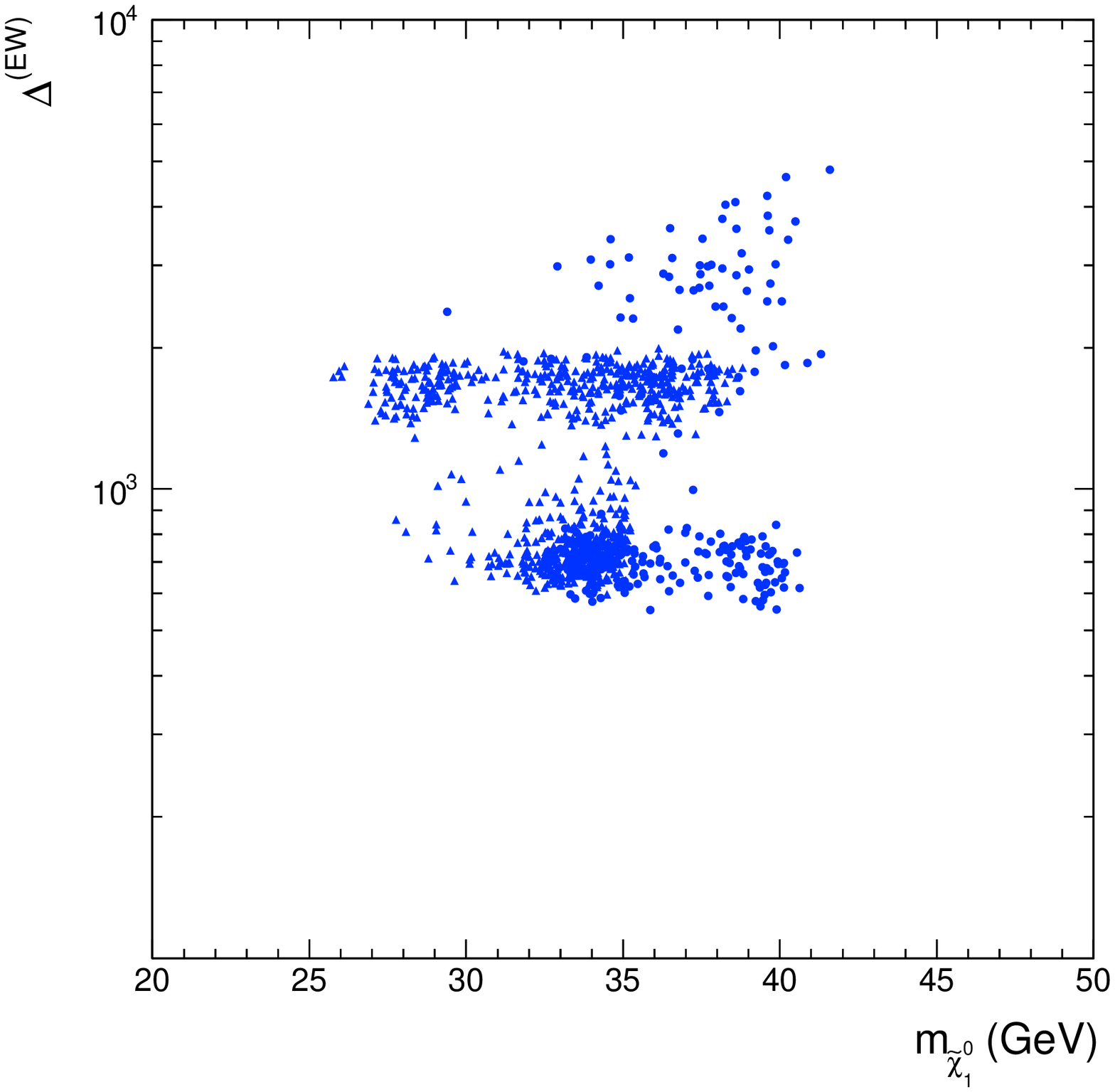,width=7.6cm}   
  \end{center}
    \captions{Electroweak fine-tuning as a function of the neutralino mass for all the experimentally allowed solutions discussed in the previous sections. 
    }
  \label{fig:EWFT}
\end{figure}

The EW fine-tuning for the points that fulfil all the experimental data are shown in Figure~\ref{fig:EWFT}. In all cases, 
the main source of fine-tuning is the $\mu$ parameter, which at the LE scale lies in the range $\sim [1500,4500]$~GeV.  
Remarkably, more than 50\% of our solutions have $\EWft$ at the percent level,\footnote{An $\EWft \gtrsim {\cal O}(100)$  
corresponds to a fine-tuning at a level $\lesssim 1\%$.} corresponding to those with lower $\mu(LE)$, since 
\begin{equation}
\Delta_{\mu}^{(\rm EW)} = \frac{\mu}{m_h^2}\frac{\partial m_h^2}{\partial \mu} 
=-4 c_\mu^2 \dfrac{\mu^2}{m_h^2} = -4 \left(\dfrac{\mu(LE)}{m_h}\right)^2 \ . 
\end{equation}

Nevertheless, in the MSSM, there are other implicit potential fine-tunings that have to be taken into account
when evaluating the global degree of naturalness.
They stem from the need of having a physical Higgs mass consistent with $\mhexp\simeq 125$~GeV and 
from the requirement of a large value of $\tan\beta$. 
Regarding the latter, moderately large values of $\tan\beta$ generically require a small value of $B\mu$ at low energy, a fact which entails a cancellation between the initial value and the radiative contribution 
from the RGE running. 
Using Eq.~(6.6) of Ref.~\cite{Casas:2014eca}, we have verified that there is no fine-tuning associated with the values of $\tan\beta$ in our allowed solutions. 

\subsection{Fine-tuning to obtain the experimental Higgs mass}

As is known, in the MSSM radiative corrections to the Higgs mass
are needed to reconcile it with the experimental value. 
A simplified expression  of such corrections \cite{Haber:1996fp, Casas:1994us,Carena:1995bx}, is
\begin{equation}\label{eq:der2}
\delta m_{h}^{2}=\frac{3 G_{F}}{\sqrt 2 \pi^{2}}m_{t}^{4}\left(\log\left(\frac{\overline m_{\tilde t}^{2}}{m_{t}^{2}}\right)+\frac{X_{t}^{2}}{\overline m_{\tilde t}^{2}}\left(1-\frac{X_{t}^{2}}{12\overline m_{\tilde t}^{2}}\right)\right)\ +\ \cdots\ ,
\end{equation}
with $\overline m_{\tilde t}$ the average stop mass and $X_{t}=A_{t}(LE)-\mu(LE) \cot\beta$.
The $X_t$-contribution that arises from the threshold corrections to the quartic coupling at the stop scale is maximised 
for $X_{t}=\sqrt 6 \overline m_{\tilde t}$ ($X_{t}\simeq  2 \overline m_{\tilde t}$ when higher orders are included). 
If this correction were not present heavy stops would be needed (of about 3~TeV once higher order corrections are included) 
for large $\tan \beta$ (and much heavier as $\tan\beta$ decreases \cite{Cabrera:2011bi,Giudice:2011cg}).
However, taking $X_t$ close to the ``maximal" value also entails a certain degree of tuning depending on how close to such value it is required to be. 
The precision needed depends, in turn, on the values of $\tan\beta$ and the stop masses \cite{Casas:2014eca}.

\begin{figure}[t!]
  \begin{center}
    \epsfig{file=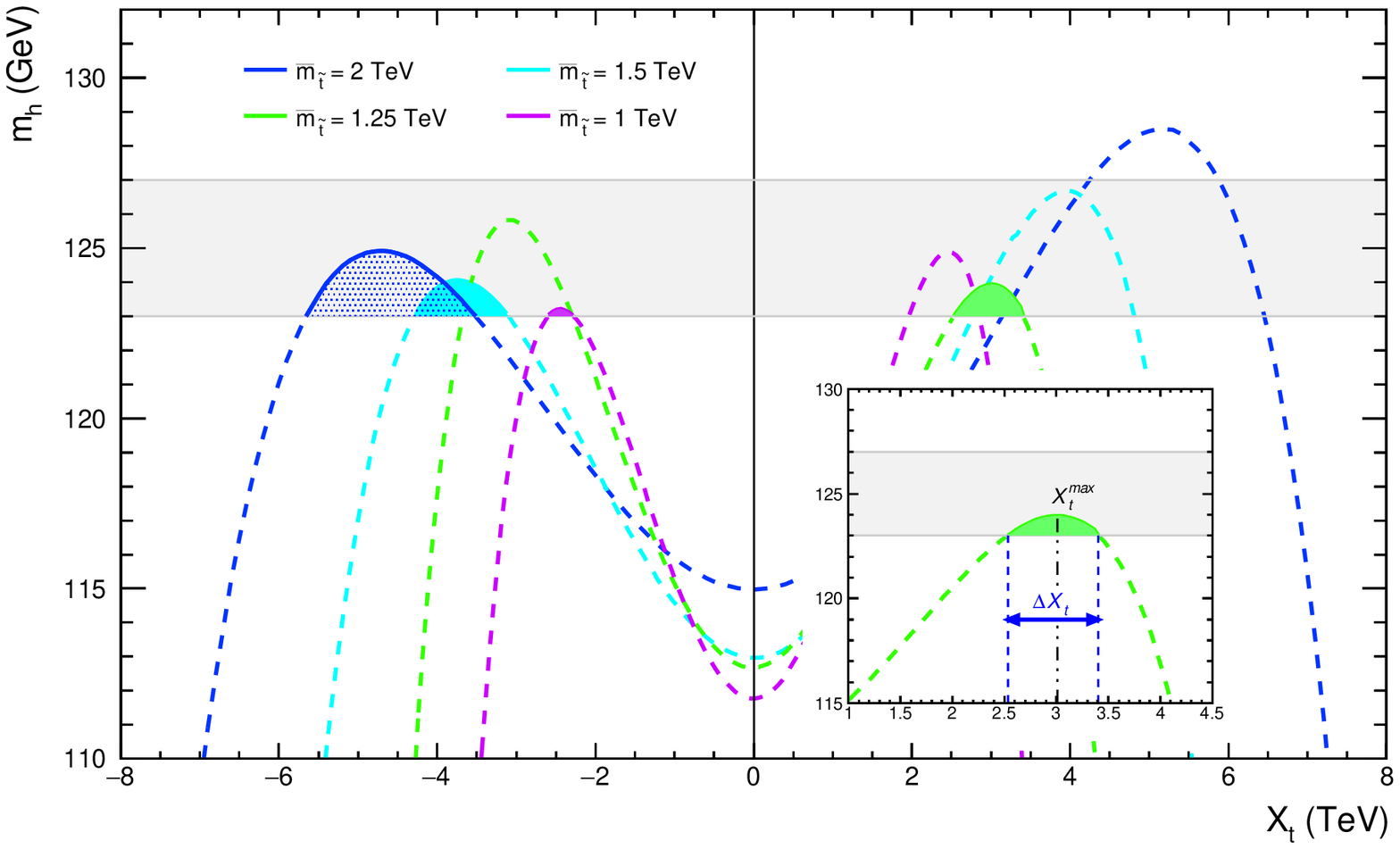,width=11.4cm}   
  \end{center}
    \captions{The Higgs mass as a function of $X_t$ for different average stop masses for four of our allowed solutions, 
    three of them with $A_U<0$ and $\mstopav = 1~\TeV$, $1.5~\TeV$ and $2~\TeV$, and one with $A_U>0$ and $\mstopav=1.25~\TeV$. 
    The grey band denotes the uncertainty on the Higgs mass, $\mhexp=125\pm2~\GeV$. 
    Solid lines and shaded regions correspond to the range of variation in $A_t(LE)$ to achieve the experimental Higgs mass. 
    Dashed lines are drawn for completeness to show that there are four possible $X_t$ values to obtain a Higgs mass within 
    the theoretical uncertainty. 
    The inset plot illustrates the criterion to determine the $p$-value measure of the fine-tuning to obtain $\mhexp$ for the solution with $A_U>0$ and $\mstopav=1.25~\TeV$. 
    }
  \label{fig:mhXt}
\end{figure}

To illustrate this potential fine-tuning, we have let $A_t(LE)$ vary freely for four of our allowed solutions with 
different average stop masses in Figure~\ref{fig:mhXt}. Three of them correspond to points of the parameter space with $A_U<0$ 
and $\mstopav = 1~\TeV$, $1.5~\TeV$ and $2~\TeV$, and the other case to $A_U>0$ and $\mstopav=1.25~\TeV$. 
The shaded regions and solid lines denote the $X_t$ range for which $m_h \geq \mhexp$. 
Dashed lines show that, in fact, there are four possible values of $X_t$ for which $m_h$ lies in the interval,  
$\mhexp=125\pm2$~GeV (the uncertainty is mainly due to the theoretical calculation)\footnote{Notice that the Higgs mass computation vary according to the code used, which could affect our 
fine-tuning estimation (see for instance \cite{Casas:2016xnl}). For this reason, we have assumed a $2~\GeV$ uncertainty on $m_h$. }.
Note that for $\mstopav\sim1$~TeV and $X_t<0$ (solid violet line), we require an almost precise value of $A_t(LE)$, around $\sim-2.5$~TeV to achieve 
$\mhexp$ (see the small shaded violet region)\footnote{Remind 
that we are in a moderately large $\tan\beta$ regime (see Figure~\ref{fig:inputs}, right-hand panel), 
then we can approximate $X_{t}=A_t(LE)-\mu(LE) \cot\beta\simeq A_t(LE)$.}. 
On the contrary, for $\mstopav\gtrsim1.25$~TeV (see shaded green, cyan and blue regions), no accurate arrangement is needed since a broad range of $X_t$ values is allowed, e.g. 
for $\mstopav\simeq1.25$~TeV, $A_t(LE)$ could take values in the range $\simeq3\pm0.5$~TeV.

The aforementioned fine-tuning is independent of that required to obtain the correct electroweak scale; therefore, if both 
tunings are present we must properly combine them \cite{Casas:2014eca}. From Figure~\ref{fig:mhXt}, we can infer that some of our solutions  
could undergo this tuning. Consequently, we have to quantify it, with a method that has a similar statistical meaning as the measure used 
for the EW fine-tuning. We have adopted the criterion described in Ref.~\cite{Casas:2014eca}, which states that 
the fine-tuning is well estimated by the $p-$value of obtaining $m_h \geq \mhexp$, i.e. 
\begin{equation}
\label{pvaluemh}
p{\rm - value} = \int_{m_h\geq \mhexp}\ d m_h\ {\cal P} (m_h) \ ,
 \end{equation}
where $ {\cal P} (m_h)$ is the probability of a Higgs mass value, given by
\begin{equation}
\label{Pmh}
{\cal P} (m_h)=\left| \frac{dX_t}{dm_h}\right| {\cal P}(X_t(m_h)) \ , 
\end{equation}
with  ${\cal P} (X_t)$ the probability distribution of $X_t$ values.\footnote{For a range of $m_h$ values, 
there are four $X_t$ values for which $m_h=\mhexp$ (see Figure~\ref{fig:mhXt}), 
so ${\cal P} (m_h)$ is the sum of four terms, corresponding to those solutions.} 
Then, the fine-tuning is evaluated as the inverse of the $p$-value. 
Note here that $X_t\simeq A_t(LE)$ is a low-energy quantity and thus a prior on it can not be defined. 
Strictly speaking, the prior should be assumed on the initial, GUT scale parameters that determine the value of $A_t(LE)$, 
(i.e. $A_U$ and $M_2$, since for the model analysed here $A_t(LE)\simeq c_{M_2}M_2+c_{A_U}A_U$), in a similar way to the previous section. 

The criterion adopted here for the the $p$-value measure is as follows. Assuming a flat prior on $A_U$ and $M_2$, 
we have obtained a range of $X_t$ values that yield different Higgs masses, similar to those of Figure~\ref{fig:mhXt}.  
Next, we have determined the $X_t$ interval, $\Delta X_t$, for which we reach the minimum Higgs mass within the $2~\GeV$ uncertainty considered throughout this work. 
Finally, we have computed the $X_t$ value that maximises $m_h$ in this interval, $X_t^{max}$ (see inset plot in Figure~\ref{fig:mhXt}), and evaluated 
the $p$-value as  
\begin{equation}
\label{pvaleAt}
p{\rm - value} =\left| \frac{\Delta X_t}{X_t^{max}}\right| \ .  
\end{equation}

\begin{figure}[t!]
  \begin{center}
    \epsfig{file=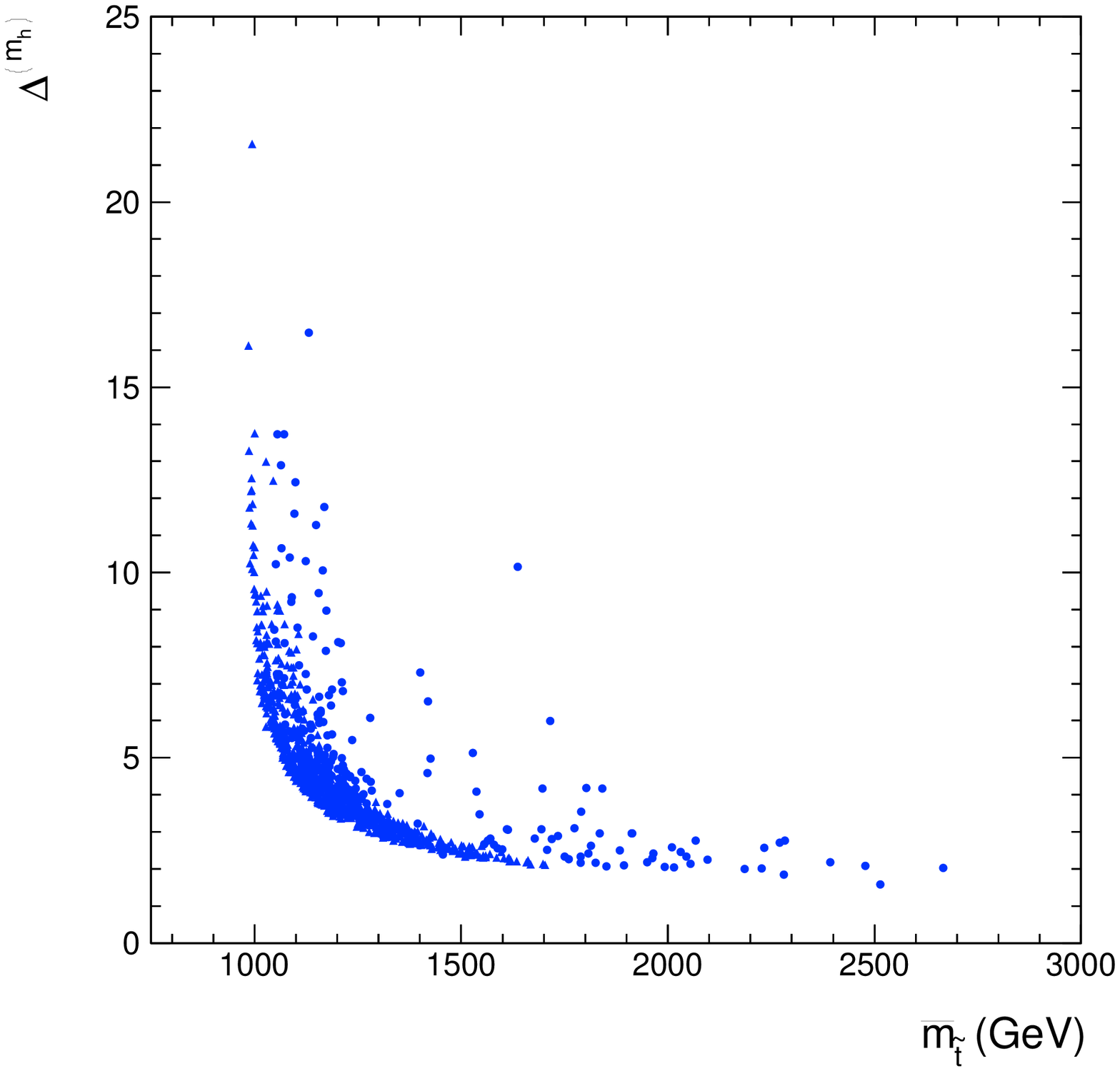,width=7.6cm}   
  \end{center}
    \captions{The fine-tuning to obtain the experimental Higgs mass as a function of the average stop mass, for the solutions 
    that fulfil all the experimental constraints.
    }
  \label{fig:mhft}
\end{figure}

Following this method, we have calculated the fine-tuning to achieve the measured Higgs mass. In Figure~\ref{fig:mhft}, 
we show our results as a function of the average stop mass. Notice that 
 for $\mstopav\gtrsim1200$~GeV, $(p{\rm-value})^{-1}$ is ${\cal O}(1)$, 
this means that in this $\mstopav$ range, and taking into account the two previously described sources of fine-tuning, 
the only important contribution is that associated with the electroweak scale.
On the other hand, for solutions with $\mstopav\lesssim1200$~GeV, even though they exhibit a mild fine-tuning to obtain $\mhexp$, 
their total fine-tuning  will grow more than one order of magnitude, when combined with $\EWft$. 
Indeed, these points correspond to solutions with $\EWft$ at the percent level, leading 
to a total fine-tuning at the per-mil level or even worse in a few cases.

\subsection{Fine-tuning to reproduce the DM relic abundance}

Besides the fine-tunings described above, there is another potential source of fine-tuning related to the
generation of the right amount of DM in the Universe, since a delicate balance between a-priori-independent quantities might be required 
in order to reproduce it. 
In such a case, we should combine this fine-tuning with that of the EW scale and, if it exists, with that to match the measured Higgs mass, 
 to select the regions of the parameter space 
that are globally less fine-tuned.

The light neutralinos analysed in this work, are bino-like, specifically they are bino-Higgsino, which requires a certain degree of 
well-tempering that could entail a fine-tuning. 
As argued in Ref.~\cite{Cabrera:2016wwr}, the ``standard measurement" of fine-tuning, which for the DM relic abundance reads
\begin{equation}
\DMft = \frac{d \log \Omega_{\rm DM}}{d \log \theta} \ ,
\end{equation}
is not appropriate for this kind of scenarios, and 
we have to evaluate the $p$-value associated with the smallness of $\Omega_{\rm DM}$.
Therefore, we can estimate $\DMft$ with the inverse of the $p$-value, i.e. for a bino-Higgsino admixture 
\begin{equation}
\label{thetapvalue}
\DMft = (p{\rm -value})^{-1}\ = \ \frac{|M_1|}{2|\mu-|M_1||}
\ . 
\end{equation}
From this expression, we conclude that there is no tuning needed, since $\mu \gg |M_1|$ and consequently $\DMft \ll 1$.

\begin{figure}[t!]
  \begin{center}
    \epsfig{file=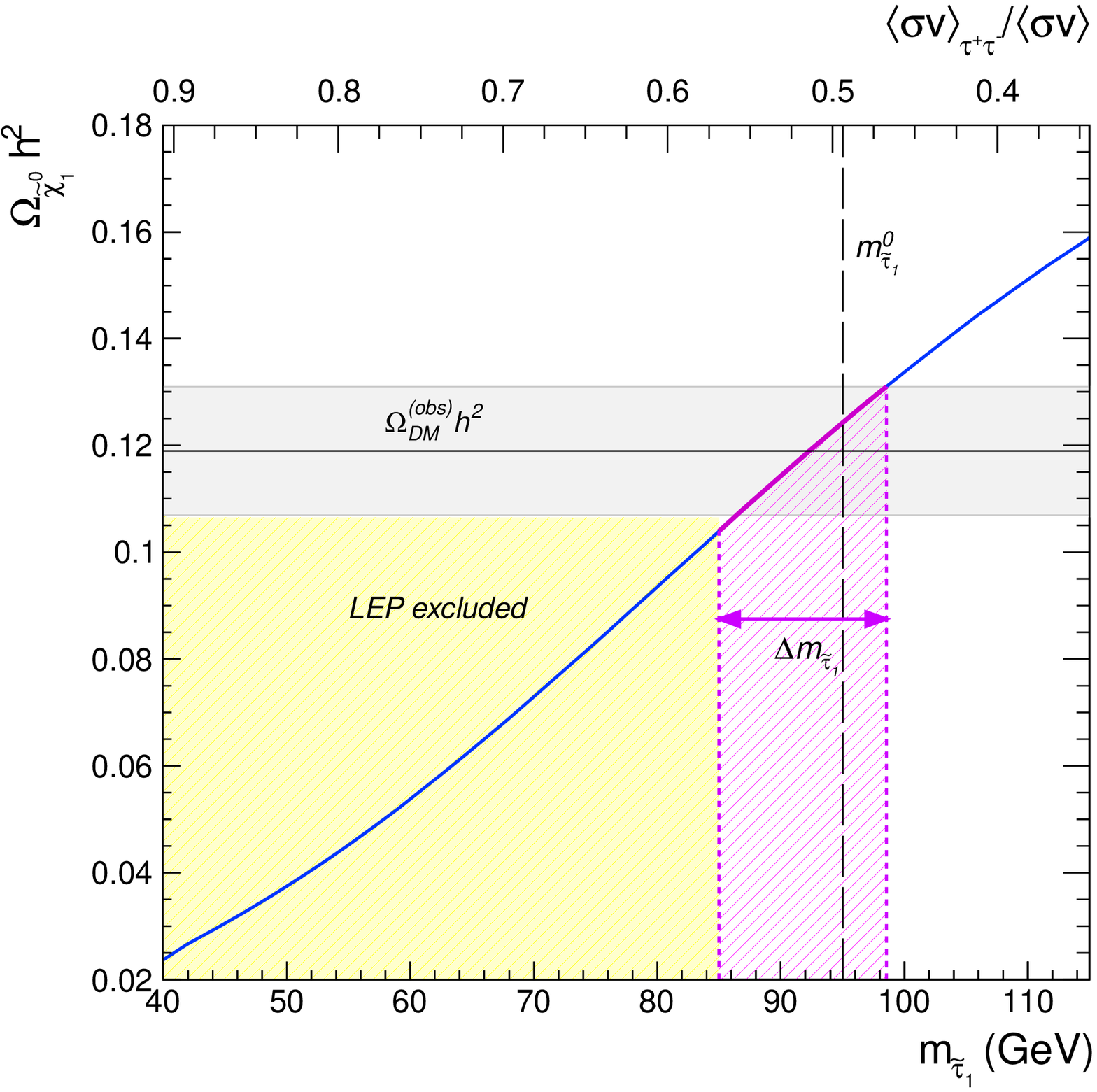,width=7.6cm}   
  \end{center}
    \captions{Relic density versus the lightest stau mass and the $p$-value criterion to estimate the fine-tuning to reproduce the DM relic density. 
    The grey band stands for $\Omega_{DM}^{(\rm obs)} h^2 = 0.119\pm0.012$ and the yellow region is excluded by the LEP bound on the stau mass.
    }
  \label{fig:relicmstau}
\end{figure}

Nonetheless, the scenarios analysed throughout this paper not only depend on the well-tempered bino-Higgsino, but as we have seen previously, 
they also rely heavily on the presence of light RH-like staus. This is due to the fact that the relic density is produced mainly by 
neutralino annihilation into $\tau^+\tau^-$ through stau exchange. The existence of this final state is controlled by $\mstauR$, 
which should be large enough to avoid the LEP limit and sufficiently small to provide the dominant annihilation channel. 
Therefore, we once again adopt a $p$-value criterion, which has been proved to be a sensible measure of $\DMft$ \cite{Cabrera:2016wwr}, 
based on this observable. For a given solution of the parameter space with the lightest stau mass, $\mstauR^0$, the $p$-value reads
\begin{equation}
\label{staupvalue}
p{\rm -value} = \ \frac{\Delta \mstauR}{\mstauR^0} \ ,
\end{equation}
where $\Delta \mstauR$ is the $\mstauR$ interval for which $\Omega_{\tilde{\chi}^0_1} h^2 \leq \Omega_{\rm DM}^{(\rm obs)} h^2$ and 
the main annihilation final state in the early Universe is $\tau^+\tau^-$. 
Figure~\ref{fig:relicmstau} highlights the use of this criterion for one of our allowed solutions. We can observe how the LEP bound shrinks significantly the value of $\Delta \mstauR$, 
which otherwise would lead to a mild fine-tuning. The upper limit on the stau mass is given by requiring the correct relic abundance and  
the channel mediated by $\tilde{\tau}_1$ being the dominant annihilation final state.

To apply this measure to our data, we have to assume a flat prior on the initial GUT scale parameters,   
that play a role in the determination of $\mstauR$, and $\tan\beta$
in a similar way as in the previous fine-tuning computations. Accordingly, the parameters that should be selected are those that 
will determine $m_{E_3}^2(LE)$, $m_{L_3}^2(LE)$, $A_\tau(LE)$ and $\mu(LE)$.  
To be more concrete, the subset of initial parameters we should considered for this calculation are\footnote{Note that 
$m_{L_3}^2(LE)\simeq c_{m_{L_3}^2}m_{L_3}^2+c_{M_2^2}M_2^2$, $m_{E_3}^2(LE)\simeq c_{m_{E_3}^2}m_{E_3}^2+c_{M_1^2}M_1^2$ and 
$A_\tau(LE) \simeq c_{A_E}A_E + c_{M_2} M_2 + c_{M_1}M_1$.}
\begin{equation}
\label{eq:DMftparam1}
\Theta_i=\left\{\mu, M_1,M_2,A_{E}, m_{L_3}^2, m_{E_3}^2, \tan\beta\right\}.
\end{equation}

Let us at this point to comment on the subtleties that we must take into account to proceed correctly. 
On the one hand, notice that $\mu$ is not a free parameter of our scan but it is the result of the REWSB, for this reason we need to enlarge the set of $\theta_i$ parameters in Eq.\eqref{eq:DMftparam1} to consider 
those that will impact on the value of $\mu(LE)$. It is worth mentioning that in the SUGRA scenarios investigated here, $M_2$ will affect both $\mu(LE)$ and $m_{L_3}^2(LE)$. 
On the other hand, when computing the fine-tuning in a quantity, say $\Omega_{\rm DM}$, without taking into account all the potential constraints 
(in this case that of the EW scale, being $G(\theta_i)=0$ the EW condition), 
the constrained $\DMft$ must be calculated projecting the unconstrained quantity $\vec \Delta^{(\rm DM)}$ into the subspace orthogonal 
to the $G(\theta_i)=0$ hypersurface in the $\{\log \theta_i\}$ space. In other words, we have to re-define the $\vec \Delta^{(\rm DM)}$ as 
\begin{equation}
\vec \Delta^{(\rm DM)}\rightarrow \vec \Delta^{(\rm DM)}-\frac{1}{|\vec \Delta G|^2}(\vec \Delta^{(\rm DM)}\cdot \vec \Delta G)\vec \Delta G\ , 
\end{equation}
where $\overrightarrow {\Delta G}\equiv \{\partial G/\partial \log\theta_i\}\propto \vec \Delta^{(\rm EW)}$. 
This issue was first noted in Ref.~\cite{Casas:2005ev}, and subsequently in Refs.~ \cite{Fichet:2012sn, Cheung:2012qy, Cabrera:2016wwr}. 
In light of this, we have calculated the $p$-value subject to there exist electroweak symmetry breaking and a valid MSSM mass spectrum 
for the following parameters, 
\begin{equation}
\label{eq:DMftparam2}
\Theta_i=\left\{M_1,M_2,A_{E}, m_{L_3}^2, m_{E_3}^2, m_{H_u}^2, m_{H_d}^2, A_U, m_{Q}^2, \tan\beta\right\}.
\end{equation}

We have evaluated then the fine-tuning to reproduce the DM relic abundance, 
$\DMft$,  as follows
\begin{equation}
\DMft_{\theta_i} = (p{\rm -value})^{-1}_{\theta_i}\ , \qquad \DMft \equiv {\rm Max}\ \left|\DMft_{\theta_i}\right|
\ . 
\end{equation}

\begin{figure}[t!]
  \begin{center}
    \epsfig{file=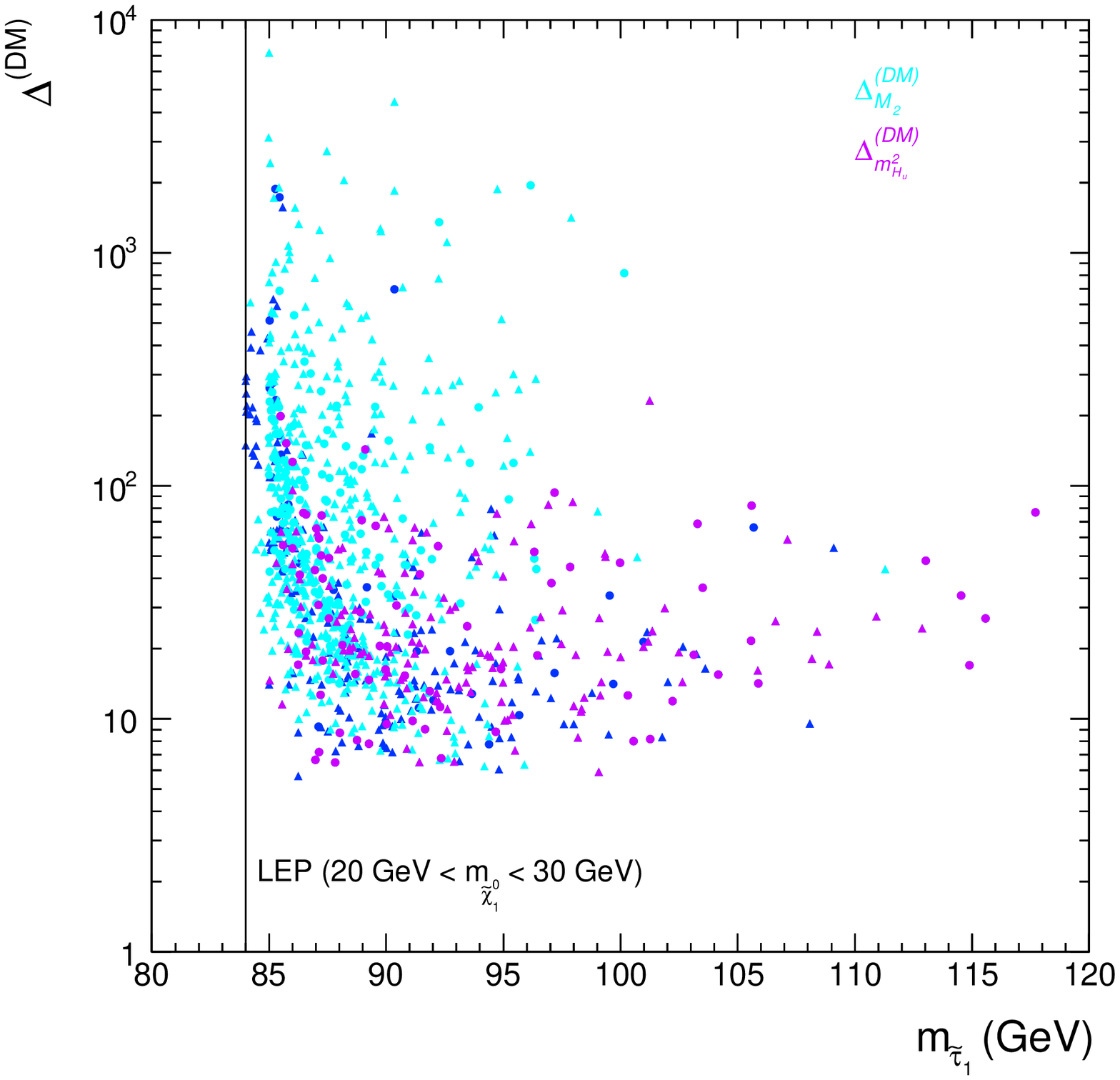,width=7.6cm}
  \end{center}
    \captions{Fine-tuning to obtain $\Omega_{\tilde{\chi}^0_1} h^2 \leq \Omega_{\rm DM}^{(\rm obs)} h^2$  as a function of $\mstauR$ for 
    the solutions that fulfil all the experimental constrains. The black solid line denotes the LEP bound for 
    $20~\GeV < m_{\tilde{\chi}^{0}_{1}} < 30~\GeV$. 
    Cyan points represent solutions with $\DMft=\DMft_{M_2}$, violet points depict solutions with $\DMft=\DMft_{m_{H_u}^2}$, 
    whereas solutions with other main sources of $\DMft$ are displayed in blue. 
    }
  \label{fig:DMft}
\end{figure}

In Figure~\ref{fig:DMft}, we present our results for $\DMft$ as a function of $\mstauR$. 
Note that the DM fine-tuning spans several orders of magnitude. 
There are a few points with  $(p{\rm-value})^{-1}\lesssim10$, this is, with no significant DM tuning.
As expected, the vast majority of solutions have a $\DMft$ above the sub-percent level or even as severe as the per-mil level. 
This is caused by the LEP limit that bounds $\mstauR$ from below depending on the neutralino mass, and hence 
shrinking considerably the $\mstauR$ interval for which $\Omega_{\tilde{\chi}^0_1} h^2 \leq \Omega_{\rm DM}^{(\rm obs)} h^2$, 
as mentioned before. 

We observe that, for a given $\mstauR$, $\DMft$ can take a broad range of values as well. This occurs because we are calculating $\DMft$ 
subject to fulfil all the potential constraints. In consequence, when any $\theta_i$ parameter  
in Eq.~\eqref{eq:DMftparam2} is varied to determine $\Delta\mstauR$, 
we must ensure, at each step as mentioned before, that there exist electroweak symmetry breaking and a valid MSSM mass spectrum. 
This is well highlighted by the fact that the main source of $\DMft$ is $M_2$ (see cyan points in Figure~\ref{fig:DMft}), 
which turns out to be the leading contribution to the running of $m_{H_u}^2$  
and the squark soft masses, and the next-to-leading contribution to the running of the left-handed slepton soft masses.\footnote{We remind the reader that we have considered $M_2=M_3$.} As a result, $M_2$ 
is the main parameter to constrain $\Delta\mstauR$ through the EW symmetry breaking condition and a valid physical mass spectrum, 
aside from the LEP limit and the DM relic density.
Other parameters that have a significant impact on $\DMft$ are $m_{H_u}^2$ (violet points), $A_U$, $m_{Q}^2$ and $M_1$(blue points).

\subsection{Total estimation of the fine-tuning}

\begin{figure}[t!]
  \begin{center}
    \epsfig{file=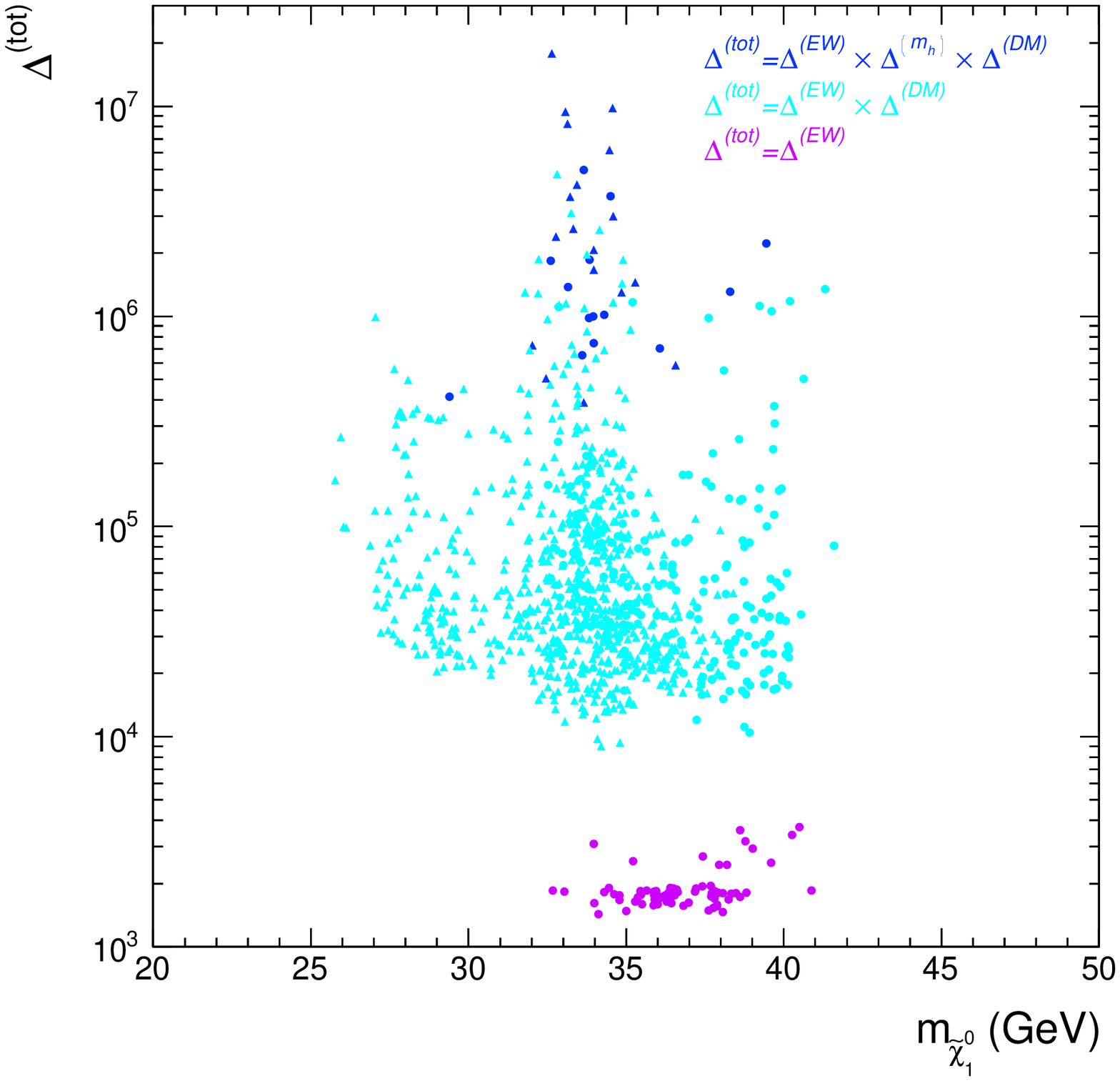,width=7.6cm}  
  \end{center}
    \captions{Total fine-tuning as a function of the lightest neutralino mass for the allowed solutions considering all the relevant sources of fine-tuning. 
    Solutions that suffer from the three fine-tunings analysed here are shown in blue ($\totft=\EWft\times\Delta^{(m_h)}\times\DMft$), 
    points that exhibit only EW and DM tunings are depicted in cyan ($\totft=\EWft\times\DMft$) and finally solutions  with only
    $\EWft$ are shown in violet. 
    }
  \label{fig:totalft}
\end{figure}

Even if $\DMft$ alone could seem less severe than $\EWft$ for most of our solutions, we should combine $\DMft$ with the 
tunings  calculated in the previous sections in order to obtain a total fine-tuning measure. 
Due to their common statistical interpretation (as $p$-values), these quantities should
be multiplicatively combined. As already pointed out, we have computed $\DMft$ considering that all constraints are fulfilled, so that 
we are allowed to directly combine $\DMft$ with the other fine-tunings without any redefinition of this quantity.

In Figure~\ref{fig:totalft}, we show the total estimation of the fine-tuning, $\totft$, for the solutions that 
fulfil all the experimental constraints. Points in blue undergo the three kinds of fine-tuning computed along this section, 
points in cyan have $\totft=\EWft\times\DMft$, while those in violet only suffer from $\EWft$. 
The latter are the least fine-tuned of our solutions, with $\totft$ at the per-mil level.

Summarising, we have addressed the naturalness issue taking into account all the potential sources of fine-tuning in the MSSM. 
We have found that the electroweak fine-tuning and that associated with the DM relic abundance are the main factors that contribute 
to the total fine-tuning of the model analysed here. 
Adopting a $p$-value measure to quantify all the different kinds of fine-tuning, 
it turns out that our results are severely tuned, with the minimum $\totft$ of ${\cal O}(10^3)$ that corresponds to the solutions 
that only undergo electroweak fine-tuning and have neutralino masses above 32 GeV. 
Solutions featuring lighter neutralinos are even more fine-tuned with $\totft > {\cal O}(10^4)$. 
Finally, let us remark that this would have gone unnoticed, if we have 
only considered the EW fine-tuning (compare Figures~\ref{fig:EWFT} and \ref{fig:totalft}).

\section{Conclusions}
\label{sec:concl}

In this paper, we have presented the conditions in the SUSY soft terms for a general supergravity theory to contain light neutralinos ($m_{\tilde{\chi}^0_1}\lesssim40$ GeV). We have explored a 12-th dimensional parameter space in order to find these solutions while fulfilling all current experimental constraints from LHC, LEP, low energy observables, direct and indirect dark matter searches. More specifically, we have applied the bounds on the Higgs mass and couplings from the LHC, which are known to be very stringent in the MSSM, and the constraints on the squarks and slepton masses from the LHC. From LEP, we have taken into account the neutralino mass dependent lower bound on slepton and chargino masses. Finally, we have used the latest determinations on the rare decays ${\rm BR}(B_s\to \mu^+\mu^-)$ and ${\rm BR}(b\to s\gamma)$.

The spectrum of solutions found exhibit a definite pattern in the gaugino sector at the GUT scale, $4\lesssim M_2/M_1\lesssim 12$, with $M_2=M_3$, highlighting that light neutralinos require a \textit{very} non universal scenario in this sector in order to fulfil the experimental constraints. This is a consequence of the LHC bounds on squarks and sleptons and of the maximisation of the bino component of the lightest neutralino. 
The third generation trilinear, $A_U$, and $\tan\beta$ parameters are very constrained by the Higgs boson properties which gives rise to a similar conclusion to that found in the universal case. 
The pattern of the slepton soft masses is, in general, very irregular but in most cases it lies below the TeV scale. 
Finally, we have found the ratio between the two Higgs soft masses, $m_{H_u}$ and $m_{H_d}$, remarkably close to one, and hence  light neutralino scenarios do not require large deviations from universality in the Higgs sector.

The phenomenology associated with these light neutralinos is somewhat different to previous studies of the MSSM defined at the EW scale. 
The difference lies in the values of the $\mu$ parameter which, in the scenarios found here, are above the TeV scale. 
Therefore, the elastic scattering cross section (both SI and SD) of neutralinos off protons and neutrons are far below the current sensitivities of the experiments, but generally above the neutrino floor. 
On the other hand, the annihilation cross section of neutralinos in the halo, in spite of being below the thermal value for an $s$-wave annihilator, 
it is in the ballpark of the Fermi-LAT searches using the most recent Pass 8 reprocessed data from dSphs. 
The accumulation of higher exposure would be a key piece to explore these models. 
Hence, we can conclude that a hint in favour of these scenarios, regarding DM searches, would be the detection of a gamma ray source 
from dSphs, while no signal would appear in direct detection experiments, at least in the near future. 

The complementarity between the LHC and DM searches is, in these cases, remarkably important. We have shown that it would be possible to probe these scenarios with searches for direct production of sleptons and charginos at the LHC. 
Some of the solutions found in the reach of the LHC, would be really challenging for both direct and indirect detection of DM, highlighting the importance of combining the different new physics search techniques. 
All in all, one might expect that future runs could shed light on the survival, or even discovery, of the scenarios presented in this work. 

Finally, we have analysed the naturalness of the solutions allowed by the experimental data, quantifying all the possible sources of tuning in the MSSM.
First of all, we have calculated the electroweak fine-tuning which despite being controlled by the TeV scale $\mu$ parameter can be as low as the percent level. 
Nevertheless, solutions with $\mstopav$ around 1 TeV are also subject to a tuning to match the measured Higgs mass, 
which is independent of that needed to reproduce the EW scale. 
A very mild tuning when considered separately, but it should be combined with the other sources of fine-tuning. 
There is also another delicate balance required by the scenarios investigated here, that needed to obtain the correct DM relic abundance, which 
relies heavily on the lightest RH-like stau mass and that could be of the same order as the EW fine-tuning or even greater due to the LEP bound. 
When we combine all these tunings, the final picture is a severely fine-tuned scenario with the least tuned solutions at the per-mil level, i.e. ${\cal O}(0.1\%)$, 
which are those that only undergo EW fine-tuning and have neutralino masses above 32 GeV. 
Lighter neutralinos exhibit an even worse fine-tuning, $<{\cal O}(0.01\%)$. 
These findings highlight the importance of taking into account all the potential fine-tunings of a model 
when looking for regions of the parameter space with as natural as possible SUSY scenarios.

\clearpage

\noindent\textbf{Acknowledgements}\\
The authors are grateful to  L. Iba\~nez and D. Cerde\~no for useful comments, and A. Casas for enlightening discussions about naturalness. We thank the support of the Consolider-Ingenio 2010 program under grant MULTIDARK CSD2009-00064, the Spanish grant FPA2015-65929-P MINECO/FEDER UE, the Spanish MINECO ``Centro de excelencia Severo Ochoa Program" under Grant No. SEV-2012-0249,
and the European Union under the ERC Advanced Grant SPLE under contract ERC-2012-ADG-20120216-320421.
S.R.  is supported by the Campus of Excellence UAM+CSIC.


\begin{thebibliography}{99}

\bibitem{Bernabei:2003za}
  R.~Bernabei, P.~Belli, F.~Cappella, R.~Cerulli, F.~Montecchia, F.~Nozzoli, A.~Incicchitti and D.~Prosperi {\it et al.},
  Riv.\ Nuovo Cim.\  {\bf 26N1} (2003) 1
  [astro-ph/0307403].
  
  \bibitem{Bernabei:2008yi}
  R.~Bernabei {\it et al.}  [DAMA Collaboration],
  Eur.\ Phys.\ J.\ C {\bf 56} (2008) 333
  [arXiv:0804.2741 [astro-ph]].
  
  \bibitem{Aalseth:2010vx}
  C.~E.~Aalseth {\it et al.}  [CoGeNT Collaboration],
  Phys.\ Rev.\ Lett.\  {\bf 106} (2011) 131301
  [arXiv:1002.4703 [astro-ph.CO]].
  
\bibitem{Angloher:2011uu}
  G.~Angloher, M.~Bauer, I.~Bavykina, A.~Bento, C.~Bucci, C.~Ciemniak, G.~Deuter and F.~von Feilitzsch {\it et al.},
  Eur.\ Phys.\ J.\ C {\bf 72} (2012) 1971
  [arXiv:1109.0702 [astro-ph.CO]].  
  
  \bibitem{Aalseth:2011wp}
  C.~E.~Aalseth, P.~S.~Barbeau, J.~Colaresi, J.~I.~Collar, J.~Diaz Leon, J.~E.~Fast, N.~Fields and T.~W.~Hossbach {\it et al.},
  Phys.\ Rev.\ Lett.\  {\bf 107} (2011) 141301
  [arXiv:1106.0650 [astro-ph.CO]].


\bibitem{Agnese:2013rvf}
  R.~Agnese {\it et al.}  [CDMS Collaboration],
  Phys.\ Rev.\ Lett.\  {\bf 111} (2013) 251301
  [arXiv:1304.4279 [hep-ex]].

  
\bibitem{Aalseth:2014jpa}
  C.~E.~Aalseth, P.~S.~Barbeau, J.~Colaresi, J.~D.~Leon, J.~E.~Fast, T.~W.~Hossbach, A.~Knecht and M.~S.~Kos {\it et al.},
  arXiv:1401.6234 [astro-ph.CO].  
  


\bibitem{Vitale:2009hr}
  V.~Vitale {\it et al.} [Fermi-LAT Collaboration],
  arXiv:0912.3828.

\bibitem{Goodenough:2009gk}
  L.~Goodenough and D.~Hooper,
  arXiv:0910.2998.

\bibitem{Hooper:2010mq}
  D.~Hooper and L.~Goodenough,
  Phys.\ Lett.\ B {\bf 697} (2011) 412
  [arXiv:1010.2752].

\bibitem{Hooper:2011ti}
  D.~Hooper and T.~Linden,
  Phys.\ Rev.\ D {\bf 84} (2011) 123005
  [arXiv:1110.0006].

\bibitem{Abazajian:2012pn}
  K.~N.~Abazajian and M.~Kaplinghat,
  Phys.\ Rev.\ D {\bf 86} (2012) 083511,
   Erratum: [Phys.\ Rev.\ D {\bf 87} (2013) 129902]
  [arXiv:1207.6047].

\bibitem{Gordon:2013vta}
  C.~Gordon and O.~Macias,
  Phys.\ Rev.\ D {\bf 88} (2013) 8,  083521
   [Phys.\ Rev.\ D {\bf 89} (2014) 4,  049901]
  [arXiv:1306.5725 [astro-ph.HE]].

\bibitem{Macias:2013vya}
  O.~Macias and C.~Gordon,
  Phys.\ Rev.\ D {\bf 89} (2014) 6,  063515
  [arXiv:1312.6671].

\bibitem{Abazajian:2014fta}
  K.~N.~Abazajian, N.~Canac, S.~Horiuchi and M.~Kaplinghat,
  Phys.\ Rev.\ D {\bf 90} (2014) 2,  023526
  [arXiv:1402.4090].

\bibitem{Daylan:2014rsa}
  T.~Daylan, D.~P.~Finkbeiner, D.~Hooper, T.~Linden, S.~K.~N.~Portillo, N.~L.~Rodd and T.~R.~Slatyer,
  Phys.\ Dark Univ.\  {\bf 12} (2016) 1
  [arXiv:1402.6703].

\bibitem{Calore:2014xka}
  F.~Calore, I.~Cholis and C.~Weniger,
  JCAP {\bf 1503} (2015) 038
  [arXiv:1409.0042].

\bibitem{Murgia:2014}
  S.~Murgia, ``Observation of the high energy gamma-ray emission towards the Galactic center'' (2014), Talk given at the Fifth Fermi Symposium, Nagoya, 20-24 October 2014.

\bibitem{Calore:2014nla}
  F.~Calore, I.~Cholis, C.~McCabe and C.~Weniger,
  Phys.\ Rev.\ D {\bf 91} (2015) 6,  063003
  [arXiv:1411.4647].

\bibitem{Calore:2015nua}
  F.~Calore, I.~Cholis and C.~Weniger,
  arXiv:1502.02805.

\bibitem{Porter:2015uaa}
  T.~A.~Porter {\it et al.} [Fermi-LAT Collaboration],
  arXiv:1507.04688.

\bibitem{Linden:2016rcf}
  T.~Linden, N.~L.~Rodd, B.~R.~Safdi and T.~R.~Slatyer,
  arXiv:1604.01026.
  
  \bibitem{Akerib:2013tjd}
  D.~S.~Akerib {\it et al.}  [LUX Collaboration],
  Phys.\ Rev.\ Lett.\  {\bf 112} (2014) 091303
  [arXiv:1310.8214 [astro-ph.CO]].
  
\bibitem{Akerib:2015rjg}
  D.~S.~Akerib {\it et al.} [LUX Collaboration],
  Phys.\ Rev.\ Lett.\  {\bf 116} (2016) no.16,  161301
  [arXiv:1512.03506 [astro-ph.CO]].

\bibitem{Akerib:2016lao}
  D.~S.~Akerib {\it et al.} [LUX Collaboration],
  Phys.\ Rev.\ Lett.\  {\bf 116} (2016) no.16,  161302
  [arXiv:1602.03489 [hep-ex]].
  
\bibitem{Akerib:2016vxi}
  D.~S.~Akerib {\it et al.},
  arXiv:1608.07648 [astro-ph.CO].
  
\bibitem{Tan:2016zwf}
  A.~Tan {\it et al.} [PandaX-II Collaboration],
  Phys.\ Rev.\ Lett.\  {\bf 117} (2016) no.12,  121303
  [arXiv:1607.07400 [hep-ex]].
  
\bibitem{Angle:2011th}
  J.~Angle {\it et al.}  [XENON10 Collaboration],
  Phys.\ Rev.\ Lett.\  {\bf 107} (2011) 051301
   [Erratum-ibid.\  {\bf 110} (2013) 249901]
  [arXiv:1104.3088 [astro-ph.CO]].
  
\bibitem{Aprile:2012nq}
  E.~Aprile {\it et al.}  [XENON100 Collaboration],
  Phys.\ Rev.\ Lett.\  {\bf 109} (2012) 181301
  [arXiv:1207.5988 [astro-ph.CO]].
  
\bibitem{Aprile:2013doa}
  E.~Aprile {\it et al.}  [XENON100 Collaboration],
  Phys.\ Rev.\ Lett.\  {\bf 111} (2013) 2,  021301
  [arXiv:1301.6620 [astro-ph.CO]].  
  

\bibitem{Ahmed:2009zw}
  Z.~Ahmed {\it et al.}  [The CDMS-II Collaboration],
  Science {\bf 327} (2010) 1619
  [arXiv:0912.3592 [astro-ph.CO]].
  
  \bibitem{Agnese:2013jaa}
  R.~Agnese {\it et al.}  [SuperCDMS Collaboration],
  Phys.\ Rev.\ Lett.\  {\bf 112} (2014) 4,  041302
  [arXiv:1309.3259 [physics.ins-det]].
  
\bibitem{Felizardo:2011uw}
  M.~Felizardo, T.~Girard, T.~Morlat, A.~C.~Fernandes, F.~Giuliani, A.~R.~Ramos, J.~G.~Marques and M.~Auguste {\it et al.},
  arXiv:1106.3014 [astro-ph.CO].
  
\bibitem{Kim:2012rza}
  S.~C.~Kim, H.~Bhang, J.~H.~Choi, W.~G.~Kang, B.~H.~Kim, H.~J.~Kim, K.~W.~Kim and S.~K.~Kim {\it et al.},
  Phys.\ Rev.\ Lett.\  {\bf 108} (2012) 181301
  [arXiv:1204.2646 [astro-ph.CO]].  
  
\bibitem{Angloher:2015ewa}
  G.~Angloher {\it et al.} [CRESST Collaboration],
  Eur.\ Phys.\ J.\ C {\bf 76} (2016) no.1,  25
  [arXiv:1509.01515 [astro-ph.CO]].
  
\bibitem{Ahmed:2011gh}
  Z.~Ahmed {\it et al.}  [CDMS and EDELWEISS Collaborations],
  Phys.\ Rev.\ D {\bf 84} (2011) 011102
  [arXiv:1105.3377 [astro-ph.CO]].
    
 \bibitem{Agnese:2014aze}
  R.~Agnese {\it et al.}  [SuperCDMS Collaboration],
  Phys.\ Rev.\ Lett.\  {\bf 112} (2014) 241302
  [arXiv:1402.7137 [hep-ex]].
  
    
  
\bibitem{Agnese:2015nto}
  R.~Agnese {\it et al.} [SuperCDMS Collaboration],
  Phys.\ Rev.\ Lett.\  {\bf 116} (2016) no.7,  071301
  [arXiv:1509.02448 [astro-ph.CO]].
  
\bibitem{Ackermann:2013yva}
  M.~Ackermann {\it et al.}  [Fermi-LAT Collaboration],
  Phys.\ Rev.\ D {\bf 89} (2014) 4,  042001
  [arXiv:1310.0828 [astro-ph.HE]].
  
\bibitem{Ackermann:2015zua}
  M.~Ackermann {\it et al.} [Fermi-LAT Collaboration],
  Phys.\ Rev.\ Lett.\  {\bf 115} (2015) no.23,  231301
  [arXiv:1503.02641 [astro-ph.HE]].
  
\bibitem{Gomez-Vargas:2013bea}
  G.~A.~G\'{o}mez-Vargas, M.~A.~S\'{a}nchez-Conde, J.~H.~Huh, M.~Peir\'{o}, F.~Prada, A.~Morselli, A.~Klypin and D.~G.~Cerde\~no {\it et al.},
  JCAP {\bf 1310} (2013) 029
  [arXiv:1308.3515 [astro-ph.HE]].
  
\bibitem{Bottino:2002ry}
  A.~Bottino, N.~Fornengo and S.~Scopel,
  Phys.\ Rev.\ D {\bf 67} (2003) 063519
  [hep-ph/0212379].
  
\bibitem{Fornengo:2010mk}
  N.~Fornengo, S.~Scopel and A.~Bottino,
  Phys.\ Rev.\  D {\bf 83} (2011) 015001
  [arXiv:1011.4743 [hep-ph]].
    
\bibitem{Vasquez:2010ru}
  D.~A.~Vasquez, G.~Belanger, C.~Boehm, A.~Pukhov and J.~Silk,
  Phys.\ Rev.\ D {\bf 82} (2010) 115027
  [arXiv:1009.4380 [hep-ph]].
  
\bibitem{Bottino:2011xv}
  A.~Bottino, N.~Fornengo and S.~Scopel,
  arXiv:1112.5666 [hep-ph].
  
  
\bibitem{Vasquez:2011yq}
  D.~A.~Vasquez, G.~Belanger and C.~Boehm,
  Phys.\ Rev.\ D {\bf 84} (2011) 095015
  [arXiv:1108.1338 [hep-ph]].
  
  \bibitem{Arbey:2012na}
  A.~Arbey, M.~Battaglia and F.~Mahmoudi,
  Eur.\ Phys.\ J.\ C {\bf 72} (2012) 2169
  [arXiv:1205.2557 [hep-ph]].
  
  \bibitem{Belanger:2012jn}
  G.~Belanger, S.~Biswas, C.~Boehm and B.~Mukhopadhyaya,
  JHEP {\bf 1212} (2012) 076
  [arXiv:1206.5404 [hep-ph]].

\bibitem{Boehm:2013gst}
  C.~Boehm, P.~S.~B.~Dev, A.~Mazumdar and E.~Pukartas,
  JHEP {\bf 1306} (2013) 113
  [arXiv:1303.5386 [hep-ph]].
  
  \bibitem{Pierce:2013rda}
  A.~Pierce, N.~R.~Shah and K.~Freese,
  arXiv:1309.7351 [hep-ph].
  
  \bibitem{Belanger:2013pna}
  G.~Bélanger, G.~Drieu La Rochelle, B.~Dumont, R.~M.~Godbole, S.~Kraml and S.~Kulkarni,
  Phys.\ Lett.\ B {\bf 726} (2013) 773
  [arXiv:1308.3735 [hep-ph]].
  
\bibitem{Scopel:2013bba}
  S.~Scopel, N.~Fornengo and A.~Bottino,
  Phys.\ Rev.\ D {\bf 88} (2013) 2,  023506
  [arXiv:1304.5353 [hep-ph]].
  
\bibitem{Hagiwara:2013qya}
  K.~Hagiwara, S.~Mukhopadhyay and J.~Nakamura,
  Phys.\ Rev.\ D {\bf 89} (2014) 015023
  [arXiv:1308.6738 [hep-ph]].  
  
\bibitem{Calibbi:2013poa}
  L.~Calibbi, J.~M.~Lindert, T.~Ota and Y.~Takanishi,
  JHEP {\bf 1310} (2013) 132
  [arXiv:1307.4119].
  
  \bibitem{Gunion:2005rw}
  J.~F.~Gunion, D.~Hooper and B.~McElrath,
  Phys.\ Rev.\ D {\bf 73} (2006) 015011
  [hep-ph/0509024].
  
\bibitem{Das:2010ww}
  D.~Das and U.~Ellwanger,
  JHEP {\bf 1009} (2010) 085
  [arXiv:1007.1151 [hep-ph]].  
  
  \bibitem{Cao:2011re}
  J.~J.~Cao, K.~i.~Hikasa, W.~Wang and J.~M.~Yang,
  Phys.\ Lett.\ B {\bf 703} (2011) 292
  [arXiv:1104.1754 [hep-ph]].
  
  \bibitem{Carena:2011jy}
  M.~Carena, N.~R.~Shah and C.~E.~M.~Wagner,
  Phys.\ Rev.\ D {\bf 85} (2012) 036003
  [arXiv:1110.4378 [hep-ph]].
  
  \bibitem{AlbornozVasquez:2011js}
  D.~Albornoz Vasquez, G.~Belanger and C.~Boehm,
  Phys.\ Rev.\ D {\bf 84} (2011) 095008
  [arXiv:1107.1614 [hep-ph]].
  
\bibitem{LopezFogliani:2012yq}
  D.~E.~Lopez-Fogliani,
  J.\ Phys.\ Conf.\ Ser.\  {\bf 384} (2012) 012014.  
  
  \bibitem{Kozaczuk:2013spa}
  J.~Kozaczuk and S.~Profumo,
  Phys.\ Rev.\ D {\bf 89} (2014) 095012
  [arXiv:1308.5705 [hep-ph]].
  
  \bibitem{Han:2014nba}
  T.~Han, Z.~Liu and S.~Su,
  JHEP {\bf 1408} (2014) 093
  [arXiv:1406.1181 [hep-ph]].
  
\bibitem{Cerdeno:2015jca}
  D.~G.~Cerdeno, M.~Peiro and S.~Robles,
  JCAP {\bf 1604} (2016) no.04,  011
  [arXiv:1507.08974 [hep-ph]].
  
  \bibitem{Cerdeno:2011qv}
  D.~G.~Cerde\~no, J.~H.~Huh, M.~Peir\'{o} and O.~Seto,
  JCAP {\bf 1111} (2011) 027
  [arXiv:1108.0978 [hep-ph]].
  
  \bibitem{Cerdeno:2014cda}
  D.~G.~Cerde\~no, M.~Peir\'{o} and S.~Robles,
  JCAP {\bf 1408} (2014) 005
  [arXiv:1404.2572 [hep-ph]].
  
\bibitem{Cerdeno:2015ega}
  D.~G.~Cerdeno, M.~Peiro and S.~Robles,
  Phys.\ Rev.\ D {\bf 91} (2015) no.12,  123530
  [arXiv:1501.01296 [hep-ph]].
  
 
  \bibitem{Boos:2002ze}
  E.~Boos, A.~Djouadi, M.~Muhlleitner and A.~Vologdin,
  Phys.\ Rev.\ D {\bf 66} (2002) 055004
  [hep-ph/0205160].

\bibitem{CMS:gya}
  [CMS Collaboration],
  CMS-PAS-HIG-12-050.  

\bibitem{Aad:2012tj}
  G.~Aad {\it et al.}  [ATLAS Collaboration],
  JHEP {\bf 1206} (2012) 039
  [arXiv:1204.2760 [hep-ex]].  
  
  \bibitem{Rosiek:1995kg}
  J.~Rosiek,
  hep-ph/9511250.
  
\bibitem{LEPSUSYWG}  
LEPSUSYWG, ALEPH, DELPHI, L3 and OPAL experiments, note LEPSUSYWG/04-01.1
(\burl{http://lepsusy.web.cern.ch/lepsusy/Welcome.html}).
  
\bibitem{Aad:2014vma}
  G.~Aad {\it et al.}  [ATLAS Collaboration],
  JHEP {\bf 1405} (2014) 071
  [arXiv:1403.5294 [hep-ex]].  
  
\bibitem{Khachatryan:2014qwa}
  V.~Khachatryan {\it et al.} [CMS Collaboration],
  Eur.\ Phys.\ J.\ C {\bf 74} (2014) no.9,  3036
  [arXiv:1405.7570 [hep-ex]].
  
\bibitem{Aaboud:2016tnv}
  M.~Aaboud {\it et al.} [ATLAS Collaboration],
  Phys.\ Rev.\ D {\bf 94} (2016) no.3,  032005
  doi:10.1103/PhysRevD.94.032005
  [arXiv:1604.07773 [hep-ex]].
  
  \bibitem{ATLAS:2017}
  [ATLAS Collaboration],
  ATLAS-CONF-2017-022.
  
  \bibitem{Aaij:2013aka}
  R.~Aaij {\it et al.}  [LHCb Collaboration],
  Phys.\ Rev.\ Lett.\  {\bf 111} (2013) 101805
  [arXiv:1307.5024 [hep-ex]].
  
  \bibitem{Chatrchyan:2013bka}
  S.~Chatrchyan {\it et al.}  [CMS Collaboration],
  Phys.\ Rev.\ Lett.\  {\bf 111} (2013) 101804
  [arXiv:1307.5025 [hep-ex]].
  
\bibitem{Galanti-talk}
{See talk by Mario Galanti, 
http://indico.cern.ch/conferenceDisplay.py?confId=264199}.  
  
\bibitem{Ciuchini:1998xy}
  M.~Ciuchini, G.~Degrassi, P.~Gambino and G.~F.~Giudice,
  Nucl.\ Phys.\ B {\bf 534} (1998) 3
  [hep-ph/9806308].  
  
  \bibitem{D'Ambrosio:2002ex}
  G.~D'Ambrosio, G.~F.~Giudice, G.~Isidori and A.~Strumia,
  Nucl.\ Phys.\ B {\bf 645} (2002) 155
  [hep-ph/0207036].
  
  \bibitem{Misiak:2006zs}
  M.~Misiak, H.~M.~Asatrian, K.~Bieri, M.~Czakon, A.~Czarnecki, T.~Ewerth, A.~Ferroglia and P.~Gambino {\it et al.},
  Phys.\ Rev.\ Lett.\  {\bf 98} (2007) 022002
  [hep-ph/0609232].
  
  \bibitem{Misiak:2006ab}
  M.~Misiak and M.~Steinhauser,
  Nucl.\ Phys.\ B {\bf 764} (2007) 62
  [hep-ph/0609241].
  
  \bibitem{Amhis:2012bh}
  Y.~Amhis {\it et al.}  [Heavy Flavor Averaging Group Collaboration],
  arXiv:1207.1158 [hep-ex].
  
\bibitem{deCarlos:1992pd}
  B.~de Carlos, J.~A.~Casas and C.~Munoz,
  Phys.\ Lett.\ B {\bf 299} (1993) 234
  [hep-ph/9211266].
  
\bibitem{Brignole:1993dj}
  A.~Brignole, L.~E.~Ibanez and C.~Munoz,
  Nucl.\ Phys.\ B {\bf 422} (1994) 125
   Erratum: [Nucl.\ Phys.\ B {\bf 436} (1995) 747]
  [hep-ph/9308271].
  
\bibitem{Baer:2000gf}
  H.~Baer, M.~A.~Diaz, P.~Quintana and X.~Tata,
  JHEP {\bf 0004} (2000) 016
  [hep-ph/0002245].
  
\bibitem{Khalil:2000ci}
  S.~Khalil, T.~Kobayashi and O.~Vives,
  Nucl.\ Phys.\ B {\bf 580} (2000) 275
  [hep-ph/0003086].
  
\bibitem{Ellis:2012aa}
  J.~Ellis and K.~A.~Olive,
  Eur.\ Phys.\ J.\ C {\bf 72} (2012) 2005
  [arXiv:1202.3262 [hep-ph]].
  
\bibitem{Baer:2012uya}
  H.~Baer, V.~Barger and A.~Mustafayev,
  JHEP {\bf 1205} (2012) 091
  [arXiv:1202.4038 [hep-ph]].
  
\bibitem{Cerdeno:2004zj}
  D.~G.~Cerdeno and C.~Munoz,
  JHEP {\bf 0410} (2004) 015
  [hep-ph/0405057].
  
\bibitem{Baek:2005wi}
  S.~Baek, D.~G.~Cerdeno, Y.~G.~Kim, P.~Ko and C.~Munoz,
  JHEP {\bf 0506} (2005) 017
  [hep-ph/0505019].
  
  \bibitem{Carena:2011aa}
  M.~Carena, S.~Gori, N.~R.~Shah and C.~E.~M.~Wagner,
  JHEP {\bf 1203} (2012) 014
  [arXiv:1112.3336 [hep-ph]].
  
  \bibitem{Feroz:2007kg}
  F.~Feroz and M.~P.~Hobson,
  Mon.\ Not.\ Roy.\ Astron.\ Soc.\  {\bf 384} (2008) 449
  [arXiv:0704.3704 [astro-ph]].
  
  \bibitem{Feroz:2008xx}
  F.~Feroz, M.~P.~Hobson and M.~Bridges,
  Mon.\ Not.\ Roy.\ Astron.\ Soc.\  {\bf 398} (2009) 1601
  [arXiv:0809.3437 [astro-ph]].
  
  \bibitem{Feroz:2013hea}
  F.~Feroz, M.~P. Hobson, E.~Cameron, and A.~N. Pettitt, {\it {Importance Nested
  Sampling and the MultiNest Algorithm}},
  \href{http://arxiv.org/abs/1306.2144}{{\tt arXiv:1306.2144}}.
  
\bibitem{Ade:2013zuv}
  P.~A.~R.~Ade {\it et al.} [Planck Collaboration],
  Astron.\ Astrophys.\  {\bf 571} (2014) A16
  doi:10.1051/0004-6361/201321591
  [arXiv:1303.5076 [astro-ph.CO]].
  
  \bibitem{Belanger:2013oya}
  G.~Belanger, F.~Boudjema, A.~Pukhov and A.~Semenov,
  Comput.\ Phys.\ Commun.\  {\bf 185} (2014) 960
  [arXiv:1305.0237 [hep-ph]].
  
\bibitem{Barducci:2016pcb}
  D.~Barducci, G.~Belanger, J.~Bernon, F.~Boudjema, J.~Da Silva, S.~Kraml, U.~Laa and A.~Pukhov,
  arXiv:1606.03834 [hep-ph].
  
\bibitem{Allanach:2001kg}
  B.~C.~Allanach,
  Comput.\ Phys.\ Commun.\  {\bf 143} (2002) 305
  [hep-ph/0104145].
  
\bibitem{Bechtle:2008jh}
  P.~Bechtle, O.~Brein, S.~Heinemeyer, G.~Weiglein and K.~E.~Williams,
  Comput.\ Phys.\ Commun.\  {\bf 181} (2010) 138
  [arXiv:0811.4169 [hep-ph]].  
  
  \bibitem{Bechtle:2013wla}
  P.~Bechtle, O.~Brein, S.~Heinemeyer, O.~Stål, T.~Stefaniak, G.~Weiglein and K.~E.~Williams,
  Eur.\ Phys.\ J.\ C {\bf 74} (2014) 2693
  [arXiv:1311.0055 [hep-ph]].
  
\bibitem{Bechtle:2013xfa}
  P.~Bechtle, S.~Heinemeyer, O.~Stål, T.~Stefaniak and G.~Weiglein,
  Eur.\ Phys.\ J.\ C {\bf 74} (2014) no.2,  2711
  [arXiv:1305.1933 [hep-ph]].
  
\bibitem{Bernon:2015hsa}
  J.~Bernon and B.~Dumont,
  Eur.\ Phys.\ J.\ C {\bf 75} (2015) no.9,  440
  [arXiv:1502.04138 [hep-ph]].
  
\bibitem{Kraml:2013mwa}
  S.~Kraml, S.~Kulkarni, U.~Laa, A.~Lessa, W.~Magerl, D.~Proschofsky-Spindler and W.~Waltenberger,
  Eur.\ Phys.\ J.\ C {\bf 74} (2014) 2868
  [arXiv:1312.4175 [hep-ph]].
  
\bibitem{Kraml:2014sna}
  S.~Kraml {\it et al.},
  arXiv:1412.1745 [hep-ph].
  
  \bibitem{Martin:1997ns}
  S.~P.~Martin,
  Adv.\ Ser.\ Direct.\ High Energy Phys.\  {\bf 21} (2010) 1
  [hep-ph/9709356].
  
\bibitem{Aparicio:2012iw}
  L.~Aparicio, D.~G.~Cerdeno and L.~E.~Ibanez,
  JHEP {\bf 1204} (2012) 126
  [arXiv:1202.0822 [hep-ph]].
  
\bibitem{Davier:2010nc}
  M.~Davier, A.~Hoecker, B.~Malaescu and Z.~Zhang,
  Eur.\ Phys.\ J.\ C {\bf 71} (2011) 1515
   Erratum: [Eur.\ Phys.\ J.\ C {\bf 72} (2012) 1874]
  [arXiv:1010.4180 [hep-ph]].
  
\bibitem{Benayoun:2012wc}
  M.~Benayoun, P.~David, L.~DelBuono and F.~Jegerlehner,
  Eur.\ Phys.\ J.\ C {\bf 73} (2013) 2453
  [arXiv:1210.7184 [hep-ph]].
  
\bibitem{Cerdeno:2013gqa}
  D.~G.~Cerdeño {\it et al.},
  JCAP {\bf 1307} (2013) 028
   Erratum: [JCAP {\bf 1309} (2013) E01]
  [arXiv:1304.1758 [hep-ph]].
  
\bibitem{Ellis:2008hf}
  J.~R.~Ellis, K.~A.~Olive and C.~Savage,
  Phys.\ Rev.\  D {\bf 77}, 065026 (2008)
  [arXiv:0801.3656 [hep-ph]].   
  
\bibitem{Cerdeno:2012ix}
  D.~G.~Cerdeno, M.~Fornasa, J.-H.~Huh and M.~Peiro,
  Phys.\ Rev.\ D {\bf 87} (2013) no.2,  023512
  [arXiv:1208.6426 [hep-ph]].
  
\bibitem{Aprile:2015uzo}
  E.~Aprile {\it et al.} [XENON Collaboration],
  JCAP {\bf 1604} (2016) no.04,  027
  [arXiv:1512.07501 [physics.ins-det]].
  
\bibitem{Cushman:2013zza}
  P.~Cushman {\it et al.},
  arXiv:1310.8327 [hep-ex].

\bibitem{Billard:2013qya}
  J.~Billard, L.~Strigari and E.~Figueroa-Feliciano,
  Phys.\ Rev.\ D {\bf 89} (2014) 023524
  [arXiv:1307.5458 [hep-ph]].
  
  \bibitem{Jungman:1995df}
  G.~Jungman, M.~Kamionkowski and K.~Griest,
  Phys.\ Rept.\  {\bf 267} (1996) 195
  [hep-ph/9506380].
  
  \bibitem{Belanger:2008sj}
  G.~Belanger, F.~Boudjema, A.~Pukhov and A.~Semenov,
  Comput.\ Phys.\ Commun.\  {\bf 180} (2009) 747
  [arXiv:0803.2360 [hep-ph]].

\bibitem{Ade:2015xua}
  P.~A.~R.~Ade {\it et al.} [Planck Collaboration],
  Astron.\ Astrophys.\  {\bf 594} (2016) A13
  [arXiv:1502.01589 [astro-ph.CO]].
   
  
\bibitem{Charles:2016pgz}
  E.~Charles {\it et al.} [Fermi-LAT Collaboration],
  Phys.\ Rept.\  {\bf 636} (2016) 1
  [arXiv:1605.02016 [astro-ph.HE]].


\bibitem{Chatrchyan:2013iqa}
  S.~Chatrchyan {\it et al.} [CMS Collaboration],
  Phys.\ Lett.\ B {\bf 733} (2014) 328
  [arXiv:1311.4937 [hep-ex]].
  
\bibitem{ATLAS:2013cma}
  [ATLAS Collaboration],
  ATLAS-CONF-2013-024.

\bibitem{ATLAS:2016uwq}
  The ATLAS collaboration [ATLAS Collaboration],
  ATLAS-CONF-2016-096.
  
\bibitem{CMS:2016gvu}
  CMS Collaboration [CMS Collaboration],
  CMS-PAS-SUS-16-024.
  
\bibitem{ATLAS:2016uzr}
  The ATLAS collaboration [ATLAS Collaboration],
  ATLAS-CONF-2016-052.
  
\bibitem{CMS:2016mwj}
  CMS Collaboration [CMS Collaboration],
  CMS-PAS-SUS-16-014.


\bibitem{CMS:2016hxa}
  CMS Collaboration [CMS Collaboration],
  CMS-PAS-SUS-16-029.
  
\bibitem{CMS:2013ega}
  CMS Collaboration [CMS Collaboration],
  CMS-PAS-FTR-13-014.
  
\bibitem{ATL-PHYS-PUB-2013-011}  
  The ATLAS collaboration [ATLAS Collaboration],
  ATL-PHYS-PUB-2013-011.
  
\bibitem{ATL-PHYS-PUB-2014-010}  
  The ATLAS collaboration [ATLAS Collaboration],
  ATL-PHYS-PUB-2014-010
  
\bibitem{Linssen:2012hp}
  L.~Linssen, A.~Miyamoto, M.~Stanitzki and H.~Weerts,
  arXiv:1202.5940 [physics.ins-det].

\bibitem{Baer:2013vqa}
  H.~Baer, M.~Berggren, J.~List, M.~M.~Nojiri, M.~Perelstein, A.~Pierce, W.~Porod and T.~Tanabe,
  arXiv:1307.5248 [hep-ph].
  
\bibitem{Moortgat-Picka:2015yla}
  G.~Moortgat-Pick {\it et al.},
  Eur.\ Phys.\ J.\ C {\bf 75} (2015) no.8,  371
  [arXiv:1504.01726 [hep-ph]].
  
  
\bibitem{CLIC:2016zwp}
  M.~J.~Boland {\it et al.} [CLIC and CLICdp Collaborations],
  arXiv:1608.07537 [physics.acc-ph].

\bibitem{Casas:2014eca}
  J.~A.~Casas, J.~M.~Moreno, S.~Robles, K.~Rolbiecki and B.~Zaldivar,
  JHEP {\bf 1506} (2015) 070
  [arXiv:1407.6966 [hep-ph]].
  
\bibitem{Ellis:1986yg}
  J.~R.~Ellis, K.~Enqvist, D.~V.~Nanopoulos and F.~Zwirner,
  Mod.\ Phys.\ Lett.\ A {\bf 1} (1986) 57.

\bibitem{Barbieri:1987fn}
  R.~Barbieri and G.~F.~Giudice,
  Nucl.\ Phys.\ B {\bf 306} (1988) 63.
  
\bibitem{Ciafaloni:1996zh}
  P.~Ciafaloni and A.~Strumia,
  Nucl.\ Phys.\ B {\bf 494} (1997) 41
  [hep-ph/9611204].
  
\bibitem{Staub:2013tta}
  F.~Staub,
  Comput.\ Phys.\ Commun.\  {\bf 185} (2014) 1773
  [arXiv:1309.7223 [hep-ph]].
 
\bibitem{Haber:1996fp}
  H.~E.~Haber, R.~Hempfling and A.~H.~Hoang,
  Z.\ Phys.\ C {\bf 75} (1997) 539
  [hep-ph/9609331].
  
\bibitem{Casas:1994us}
  J.~A.~Casas, J.~R.~Espinosa, M.~Quiros and A.~Riotto,
  Nucl.\ Phys.\ B {\bf 436} (1995) 3
   Erratum: [Nucl.\ Phys.\ B {\bf 439} (1995) 466]
  [hep-ph/9407389].
  
\bibitem{Carena:1995bx}
  M.~Carena, J.~R.~Espinosa, M.~Quiros and C.~E.~M.~Wagner,
  Phys.\ Lett.\ B {\bf 355} (1995) 209
  [hep-ph/9504316].

\bibitem{Cabrera:2011bi}
  M.~E.~Cabrera, J.~A.~Casas and A.~Delgado,
  Phys.\ Rev.\ Lett.\  {\bf 108} (2012) 021802
  [arXiv:1108.3867 [hep-ph]].
  
\bibitem{Giudice:2011cg}
  G.~F.~Giudice and A.~Strumia,
  Nucl.\ Phys.\ B {\bf 858} (2012) 63
  [arXiv:1108.6077 [hep-ph]].
  
\bibitem{Casas:2016xnl}
  J.~A.~Casas, J.~M.~Moreno, S.~Robles and K.~Rolbiecki,
  Eur.\ Phys.\ J.\ C {\bf 76} (2016) no.8,  450
  [arXiv:1602.06892 [hep-ph]].

\bibitem{Cabrera:2016wwr}
  M.~E.~Cabrera, J.~A.~Casas, A.~Delgado, S.~Robles and R.~Ruiz de Austri,
  JHEP {\bf 1608} (2016) 058
  [arXiv:1604.02102 [hep-ph]].
  
\bibitem{Casas:2005ev}
  J.~A.~Casas, J.~R.~Espinosa and I.~Hidalgo,
  JHEP {\bf 0503} (2005) 038
  [hep-ph/0502066].

\bibitem{Fichet:2012sn}
  S.~Fichet,
  Phys.\ Rev.\ D {\bf 86} (2012) 125029
  [arXiv:1204.4940 [hep-ph]].

\bibitem{Cheung:2012qy}
  C.~Cheung, L.~J.~Hall, D.~Pinner and J.~T.~Ruderman,
  JHEP {\bf 1305} (2013) 100
  [arXiv:1211.4873 [hep-ph]].

  
  
  
\end{thebibliography}
\end{document}